\documentclass[fleqn,usenatbib]{mnras}

\usepackage{newtxtext,newtxmath}

\usepackage[T1]{fontenc}
\usepackage{ae,aecompl}

\usepackage{graphicx}	
\usepackage{amsmath}	
\usepackage{amssymb}	
\usepackage[english]{babel}
\usepackage{blindtext}
\usepackage{supertabular}
\usepackage[colorinlistoftodos]{todonotes}
\usepackage{verbatim}
\usepackage{adjustbox}
\usepackage{array}
\usepackage{tablefootnote}

\title[24 lensed quasars from Gaia]{Gravitationally Lensed Quasars in \textit{Gaia}: II. Discovery of 24 Lensed Quasars}

\author[Cameron A. Lemon et al.]{Cameron A. Lemon,$^{1,2}$\thanks{E-mail: cl522@ast.cam.ac.uk}
Matthew W. Auger$^{1}$,
Richard G. McMahon$^{1, 2}$,
\newauthor
Fernanda Ostrovski$^{1, 2}$
\\
$^{1}$Institute of Astronomy, University of Cambridge, Madingley Road, Cambridge CB3 0HA, UK\\
$^{2}$Kavli  Institute  for  Cosmology,  University  of  Cambridge,  Madingley Road, Cambridge CB3 0HA, UK
}

\date{Accepted XXX. Received YYY; in original form ZZZ}

\pubyear{2017}

\begin{document}
\label{firstpage}
\pagerange{\pageref{firstpage}--\pageref{lastpage}}
\maketitle
\newcommand{\Gaia}{\textit{Gaia} }
\newlength{\widthplus}
\newcommand{\minuspalette}[2]{\settowidth{\widthplus}{$#1+$}\mathbin{\makebox[\widthplus]{$#1#2$}}}
\newcommand{\goodminus}{\mathpalette{\minuspalette}{\mathchar"2200}}
{\catcode`-13\gdef-{\goodminus}}
\mathcode`-"8000

\newcommand{\MINUS}{\kern 0.102em --\kern 0.102em }

\begin{abstract}
We report the discovery, spectroscopic confirmation and preliminary characterisation of 24 gravitationally lensed quasars identified using \Gaia observations. Candidates were selected in the Pan-STARRS footprint with quasar-like WISE colours or as photometric quasars from SDSS, requiring either multiple detections in \Gaia or a single \Gaia detection near a morphological galaxy. The Pan-STARRS \textit{grizY} images were modelled for the most promising candidates and 60 candidate systems were followed up with the William Herschel Telescope. 13 of the lenses were discovered as \Gaia multiples and 10 as single \Gaia detections near galaxies. We also discover 1 lens identified through a quasar emission line in an SDSS galaxy spectrum. The lenses have median image separation 2.13\arcsec \ and the source redshifts range from 1.06 to 3.36. 4 systems are quadruply-imaged and 20 are doubly-imaged. Deep CFHT data reveal an Einstein ring in one double system. We also report 12 quasar pairs, 10 of which have components at the same redshift and require further follow-up to rule out the lensing hypothesis. We compare the properties of these lenses and other known lenses recovered by our search method to a complete sample of simulated lenses to show the lenses we are missing are mainly those with small separations and higher source redshifts. The initial \Gaia data release only catalogues all images of $\sim$ 30\% of known bright lensed quasars, however the improved completeness of \Gaia data release 2 will help find all bright lensed quasars on the sky.
\end{abstract}


\section{Introduction}
There are currently fewer than 40 known gravitationally lensed quasars with image separations over 2$\arcsec$. The multiple images of these bright wider-separation systems can be monitored to determine time delays, which are used to calculate a time delay distance and hence infer the Hubble constant \citep{refsdal1966,bonvin2016}. This is a promising method to shed light on the apparent tension between local distance measurements \citep{riess2016} and CMB measurements \citep{planck2016}.
Lensed quasars are also powerful tools to probe the many facets of galaxy evolution: lens mass distributions \citep{schechter2004, bate2008, bate2011,mediavilla2009}, source quasar accretion discs \citep{rauch1991, pooley2007, jimenez2015}, source quasar host galaxies \citep{peng2006, ding2017, bayliss2017}, etc. However the results of all these studies are currently hindered by the dearth of known lensed quasars.

The first Pan-STARRS data release \citep{chambers2016} has provided arcsecond-resolution, multi-wavelength (\textit{grizY}) imaging of three quarters of the sky. This provides the perfect dataset to increase the number of bright lensed quasars through area alone. Furthermore the multi-wavelength data allow characterisation and pre-selection of the most promising candidates to spectroscopically follow-up through the use of residual features, goodness of fit and SED similarity. While many techniques exist for identifying lenses from purely photometric data \citep{ostrovski2016, schechter2017, williams2017}, all are plagued by contaminant systems that resemble lensed quasars. These include compact star-forming galaxy pairs, binary quasars and quasars projected close to blue stars and galaxies. Additionally, the resolution of ground-based imaging data often blends the separate components of lensed quasars into just one catalogued object. This blending by the point spread function (PSF) on the scale of the image separation leads to contaminant systems being confused for lensed quasars. One way to overcome this blending issue is to use the excellent resolution of \Gaia \citep{prusti2016}, which is able to separate two point sources separated by $\sim$ 0.1\arcsec \citep{fabricius2016}. Only a catalogue of detections is currently released from the \Gaia collaboration \citep{brown2016}, however this is enough to quickly reduce the number of contaminants. Even though \Gaia has not catalogued all components of close pairs \citep{arenou2017}, the knowledge that a point source exists in a system removes some star-forming galaxy contaminants from the search.

The selection, follow-up and modelling of lensed quasars using Pan-STARRS and \Gaia data forms the basis of this paper. In Section \ref{selection} we describe the search techniques and the starting photometric quasar catalogues. Section \ref{observation} explains the observations and their outcomes, and in Section \ref{modelling} we create light profile and mass models for the systems. Finally we discuss the individual systems and summarise our findings in Sections \ref{discussion} and \ref{conclusions}. Wide-field Infrared Survey Explorer (WISE) magnitudes are quoted in the AB system. The conversions from Vega to AB for ALLWISE data are $W1_{\text{AB}}$ = $W1_{\text{Vega}}$ + 2.699 and $W2_{\text{AB}}$ =  $W2_{\text{Vega}}$ + 3.339 which are given in \citet{jarrett2011} and in the ALLWISE explanatory  supplement. When required a flat cosmology with $\Omega_{\text{m}}$ = 0.3 and $H_0$ = 70$\text{km}\text{s}^{-1}\text{Mpc}^{-1}$ was used.

\section{Lens Candidate Selection} \label{selection}
The following section describes the candidate selection process. The declination range was limited to $\gtrsim -30$, i.e. requiring \textit{grizY} imaging from Pan-STARRS. For the majority of searches we required the systems to have a galactic latitude, $|b|>15$, however in some right ascension ranges (RA$\sim$80, along the galactic anti-centre) this was relaxed. Before a final sample was established, the local \Gaia stellar density for each candidate was required to be less than 50000 stars per square degree (calculated by counting \Gaia sources within a 100$\arcsec$ radius), in order to remove star clusters.

Two quasar candidate catalogues are created, and two \textit{Gaia}-based selection methods are applied to these catalogues to generate our final sample, as described in the following subsections.

\subsection{Photometric quasar candidate catalogues} \label{catalogues}

\subsubsection{I. WISE mid-infrared colours} \label{wisecolours}

\citet{stern2012} have shown that the W1 and W2 bands of WISE \citep{wright2010} can be used to select AGN by applying the colour criterion of W1$-$W2$\geq$0.16 (AB). One advantage of this selection technique is its simplicity and effectiveness \citep{wu2012,schechter2017}, however a downside for our purposes is that lensed quasar photometry can be strongly affected by the lensing galaxy, leading to WISE colours bluer than those of isolated quasars.

To overcome this we apply a looser WISE criterion of W1$-$W2$\geq-$0.14. With this limit we do not expect an unreasonable number of contaminants still meeting our \Gaia detection criteria of Section \ref{gaiamorphology}. The main contaminant created by this lower limit is quasar+star projections.

We ensure that the WISE detections are robust in W1 and W2 by requiring catalogue uncertainties and a W1 value brighter than 18.2. After matching to Pan-STARRS and keeping objects with $i$-band PSF magnitudes brighter than 21, our initial WISE-selected quasar candidate list has 1298877 objects with $|b|>15$.

\subsubsection{II. SDSS GMM photometric quasars}
Recent papers \citep[e.g.][ Ostrovski in prep.]{williams2017, agnello2018} have shown that the Sloan Digital Sky Survey (SDSS) imaging data still contain bright lensed quasars, which had not been targetted for spectroscopy and hence were missed by the SDSS Quasar Lens Search \citep[SQLS,][]{oguri2006}.

Since the SDSS imaging dataset includes $u$-band data---which are particularly useful for selecting AGN---we create a complementary catalogue to the WISE selection above by applying Gaussian mixture modelling (GMM) classification to SDSS objects as in \citet{ostrovski2017}. This is a morphology-independent selection based on $u-g$, $g-i$ and $i-$W1 colours. Classification is divided into four classes: stars, galaxies, and low- and high-redshift quasars ($z\lesssim2.7$ and $z\gtrsim2.7$ respectively). This classification is applied to all SDSS objects with $psfMag\_i<21$. Our final GMM quasar candidate catalogue is composed of all objects that have a combined (low and high redshift) quasar probability > 0.5. This results in 1158557 quasar candidates.

\subsection{Morphology Selection} \label{gaiamorphology}
Once a set of photometric quasar candidates is selected, we attempt to remove the objects that are not lensed quasars---mainly isolated quasars or misclassified star-forming galaxies. To this end we use \Gaia data, which has a spatial resolution of $\sim0\farcs1$ \citep{fabricius2016}. \Gaia is built to detect stars in our galaxy, but naturally detects bright quasars as a by-product. Therefore we search our quasar candidate catalogues for objects in which multiple \Gaia sources are detected, as would be expected for multiply imaged quasars. While this does not remove quasar+star projections from our candidates, it removes many star-forming galaxies and isolated quasars, since at most one \Gaia detection is expected for these contaminants.

Since \Gaia does not reliably detect all images of most lensed quasars, we also describe a simple morphology selection using just one \Gaia detection \citep{lemon2017}. While this selection naturally removes fewer contaminants, it is able to recover 45 known lenses. The details of the two methods are described below. They are both applied to each of the quasar catalogues described in \ref{catalogues}; Table \ref{tab:candidates} shows the number of candidates each technique and quasar catalogue produced, given the selection criteria. Note that the numbers are not exclusive.

\subsubsection{Multiple \Gaia detections} \label{multigaia}
Our first selection technique is to find quasar candidates with multiple \Gaia detections. We require at least two \Gaia detections within $4$\arcsec \ of each other. Lensed quasars with separations above this are very rare \citep[e.g.][]{oguri2010} and the number of contaminant systems increases roughly proportionally to the maximum image separation allowed in a search. Furthermore \Gaia is most useful at combating the blending of smaller-separation systems in ground-based optical survey data, in which lenses with images separated by more than 4\arcsec \ should already be deblended. This technique is applied in a two-step process: firstly all quasar candidates are matched to \Gaia within 2\arcsec \ of the Pan-STARRS detection and secondly this is matched to \Gaia again within 4\arcsec \ of the initial \Gaia position. We choose the first matching distance to be only 2\arcsec \ since this retains all known lenses with two \Gaia detections while removing some single quasar candidates projected near stars with \Gaia detections.

\subsubsection{\Gaia detection near morphological galaxy} \label{singlegaiamorph}
Cross-matching \Gaia to known lensed quasars \citep{lemon2017} demonstrates that only 1 in 5 small-separation lensed quasars have all quasar images detected by \Gaia, even when all images should be detected. Though this fraction increases with separation \citep[one third at the largest separations;][]{agnellogaia}, requiring multiple \Gaia detections will miss the majority of lenses. One way to find some of these ``missing" lenses is to perform a search depending on only one \Gaia detection. We do this but require a morphological galaxy within 4\arcsec \ of the quasar candidate, removing contamination from wider-separation star+quasar projections. If the single \Gaia detection is indeed a quasar, a bright galaxy within 4\arcsec \ is a strong candidate for acting as a foreground lens. We crossmatch our quasar candidates to \Gaia within 4\arcsec \ and then back to Pan-STARRS within 4\arcsec \ of our original quasar candidate but requiring the criterion of $r_{PSF}-r_{KRON} > 0.2$ for the new match. This extended object can be the original quasar candidate.

\subsection{Final Lens Candidate Catalogue}
After applying the \Gaia multiple and single detection techniques to the two quasar candidate catalogues, we apply two further filters. The first is requiring the astrometric excess noise \citep[AEN, ][]{lindegren2012,lindegren2016} for each \Gaia detection to be less than 10~mas. The AEN is a useful indicator for point source/galaxy separation \citep{belokurov2016, koposov2017} which holds for known lensed quasars \citep{lemon2017}. Therefore we can remove many star-forming galaxies from our search by applying the simple cut AEN$<10$~mas. The second filter is removing all candidates with local \Gaia stellar densities above 50000 stars per square degree. This is calculated by counting all \Gaia detections within 100$\arcsec$.

The catalogues are then stacked and duplicates from the two quasar candidate selection techniques are removed, leaving 109941 \Gaia singles and 31486 \Gaia pairs. After selecting $\sim$200 of the most promising candidates through visual inspection, the Pan-STARRS \textit{grizY} images are modelled simultaneously as described in Section \ref{pixelmodelling}. This is to ensure the postulated quasar images have similar colours and to prioritise systems with residual features consistent with a lensing galaxy.

\newcolumntype{R}[2]{%
    >{\adjustbox{angle=#1,lap=\width-(#2)}\bgroup}%
    l%
    <{\egroup}%
}
\newcommand*\rot{\multicolumn{1}{R{90}{1em}}}
\begin{table}
	\centering
	\caption{Candidate numbers for the two quasar catalogues and the two \Gaia morphological selection methods.}
	\label{tab:candidates}
	\begin{tabular}{clc}
		\hline
        \hline
        &\textbf{WISE quasars (W1$-$W2$>$-0.14)} & 1298877\\
        \hline
        \hline
		\noalign{\vskip 0.5mm}    
        &Gaia matches < 2$\arcsec$ & 416990\\
		\noalign{\vskip 0.5mm} 
        &2 Gaia matches within 4$\arcsec$ of each other & 9125\\
		\noalign{\vskip 0.5mm} 
        & Gaia AEN < 10mas & 8889\\
		\noalign{\vskip 0.5mm} 
        \rot{\rlap{\tiny{MULTIPLES}}} & Stellar density < 50000 / sq. degree & \textbf{8447}\\
        \noalign{\vskip 0.5mm} 
        \hline
        \noalign{\vskip 0.5mm} 
        &Gaia matches <4$\arcsec$ & 428559\\
        \noalign{\vskip 0.5mm} 
        &Gaia singles with $r_{PSF}-r_{KRON}$ > 0.2 & 120817\\
        \noalign{\vskip 0.5mm} 
        &Gaia AEN < 10mas & 80595\\
        \noalign{\vskip 0.5mm} 
        \rot{\rlap{\tiny{\ \ SINGLES}}} &Stellar density < 50000 / sq. degree & \textbf{80206}\\
        \noalign{\vskip 0.5mm} 
		\hline
        \noalign{\vskip 2.5mm} 
        \hline
        \hline
        &\textbf{SDSS GMM quasars} & 1158557 \\
        \hline
        \hline
        \noalign{\vskip 0.5mm} 
        &Gaia matches < 2$\arcsec$ & 686311\\
        \noalign{\vskip 0.5mm} 
        &2 Gaia matches within 4$\arcsec$ of each other & 24851\\
        \noalign{\vskip 0.5mm} 
        &Gaia AEN < 10mas & 24765\\
        \noalign{\vskip 0.5mm} 
        \rot{\rlap{\tiny{MULTIPLES}}} &Stellar density < 50000 / sq. degree & \textbf{24749}\\
        \noalign{\vskip 0.5mm} 
        \hline
        \noalign{\vskip 0.5mm} 
        &Gaia matches < 4$\arcsec$ & 710052\\
        \noalign{\vskip 0.5mm} 
        &Gaia singles with $r_{PSF}-r_{KRON}$ > 0.2 & 54352\\
        \noalign{\vskip 0.5mm} 
        &Gaia AEN < 10mas & 50491\\
        \noalign{\vskip 0.5mm} 
        \rot{\rlap{\tiny{\ \ SINGLES}}} &Stellar density < 50000 / sq. degree & \textbf{50488}\\
		\hline
	\end{tabular}
\end{table}

\section{Observations} \label{observation}
Spectra of 60 candidates were taken with the Intermediate dispersion Spectrograph and Imaging System (ISIS) on the 4.2-m William Herschel Telescope (WHT) on the nights of 31 March, 1 April, 12 and 13 September 2017. Since we only needed to identify broad emission line features, we used the low resolution gratings, R158 (121 \AA  mm$^{-1}$/ 1.81 \AA  pixel$^{-1}$) for the red arm and B300 (64 \AA  mm$^{-1}$/ 0.86 \AA  pixel$^{-1}$) for the blue arm. Each lens candidate was positioned along a 1$\arcsec$-wide slit to capture both quasar images. Multiple position angles were used for one quad candidate, J1721+8842.

After masking cosmic rays and subtracting the sky background, the spectra were visually inspected for broad emission lines in the separated peaks and 1-D spectra were extracted using Gaussian apertures with 0.5\arcsec \ width.

\subsection{Results}
Table \ref{tab:observations} shows a summary of the observations with sky positions, candidate selection method, outcome of the observation and WISE and \Gaia magnitudes.

\begin{figure*}
	\includegraphics[width=\textwidth]{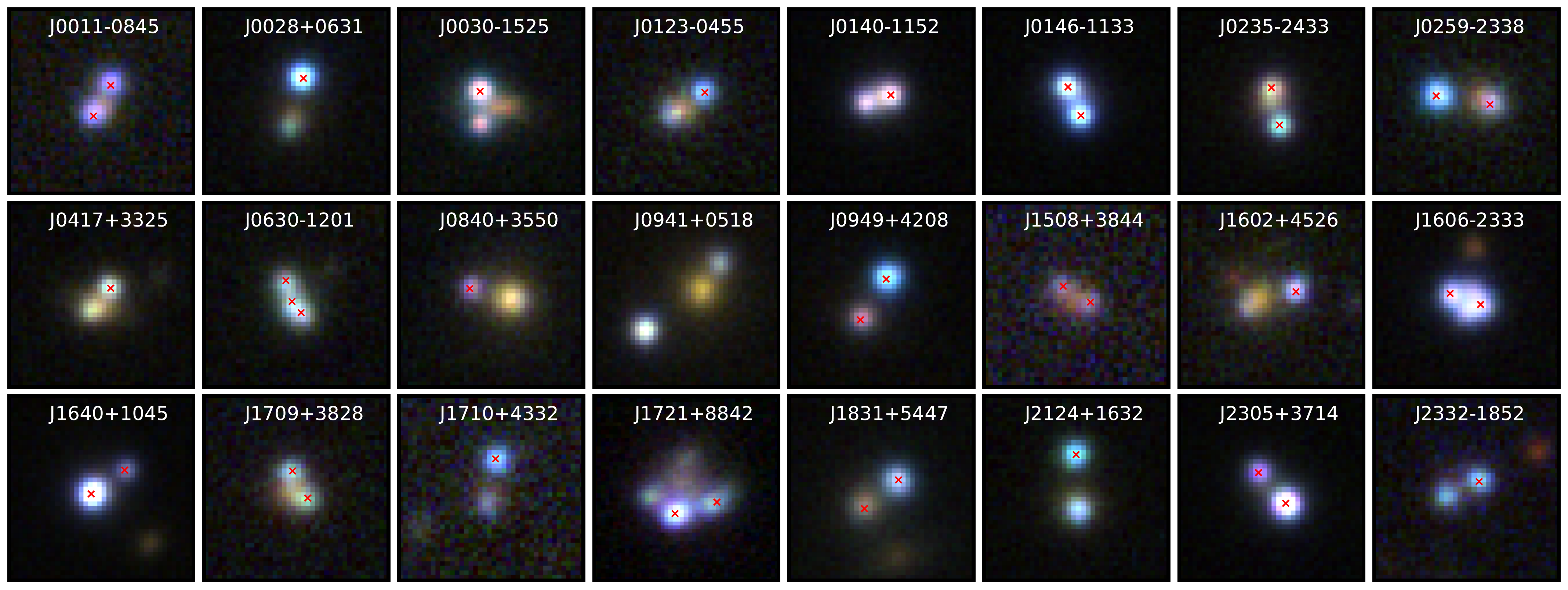}
    \caption{Pan-STARRS \textit{gri} colour images of the confirmed lenses with \Gaia detections overlaid with red crosses. Cutouts are 10$\arcsec$ on the side.}
    \label{fig:overlaid_gri}
\end{figure*}

\begin{center}
\begin{figure*}
	\includegraphics[width=\textwidth]{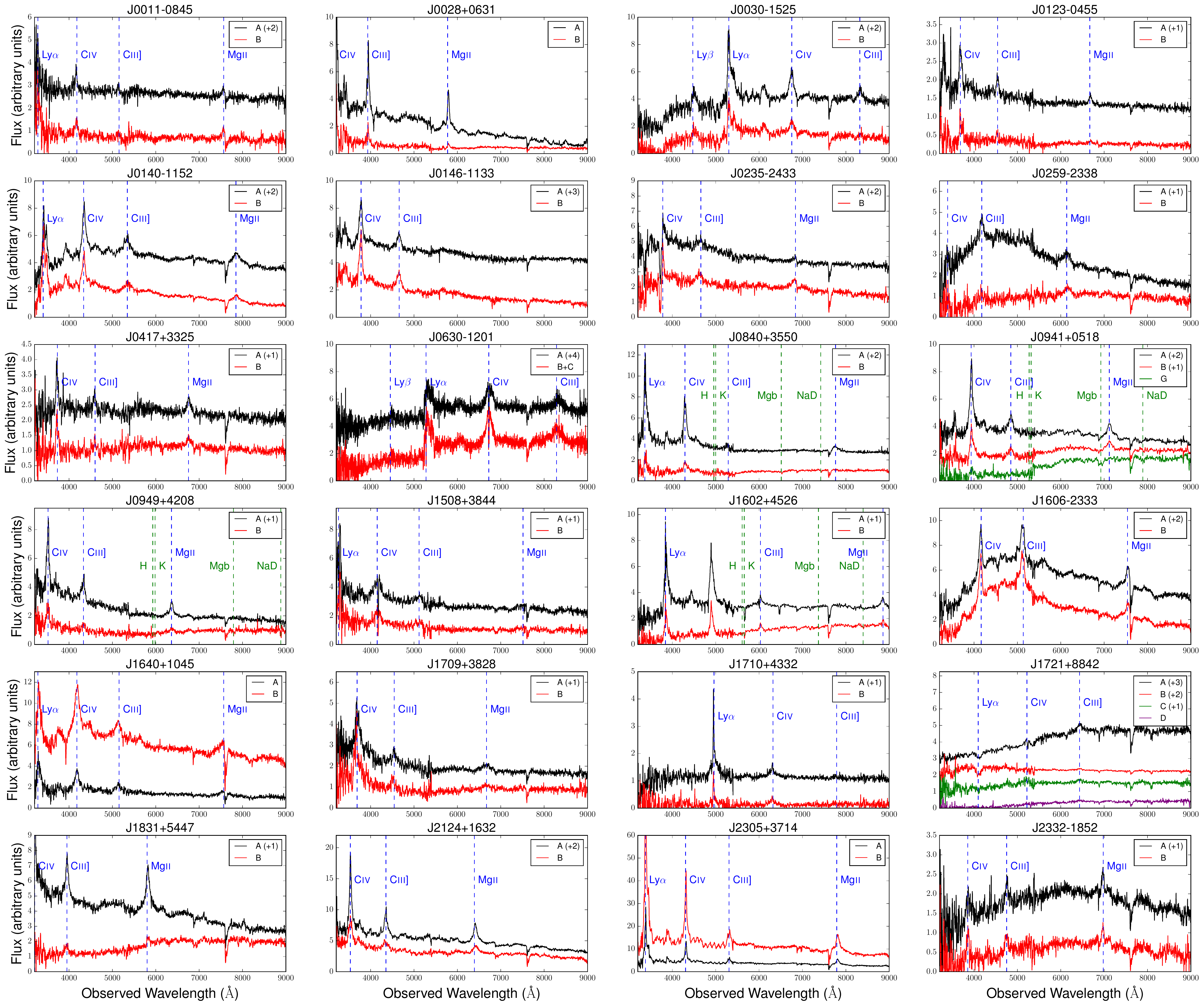}
    \caption{Spectra of the confirmed gravitationally lensed quasars. Quasar emission lines are labelled in blue, and lens galaxy absorption lines in green where identified.}
    \label{fig:spectra}
\end{figure*}
\end{center}

We have classified 24 objects as lensed quasars since the spectra reveal the presence of (at least) two quasars at the same redshift and the pixel modelling of the Pan-STARRS images reveals a lensing galaxy. Figure \ref{fig:overlaid_gri} shows Pan-STARRS \textit{gri} cutouts of the confirmed lens systems with \Gaia detections overlaid. Figure \ref{fig:spectra} shows the component spectra for each lens. We are able to establish the lens galaxy redshift for 4 lenses.

\begin{figure}
	\includegraphics[width=\columnwidth]{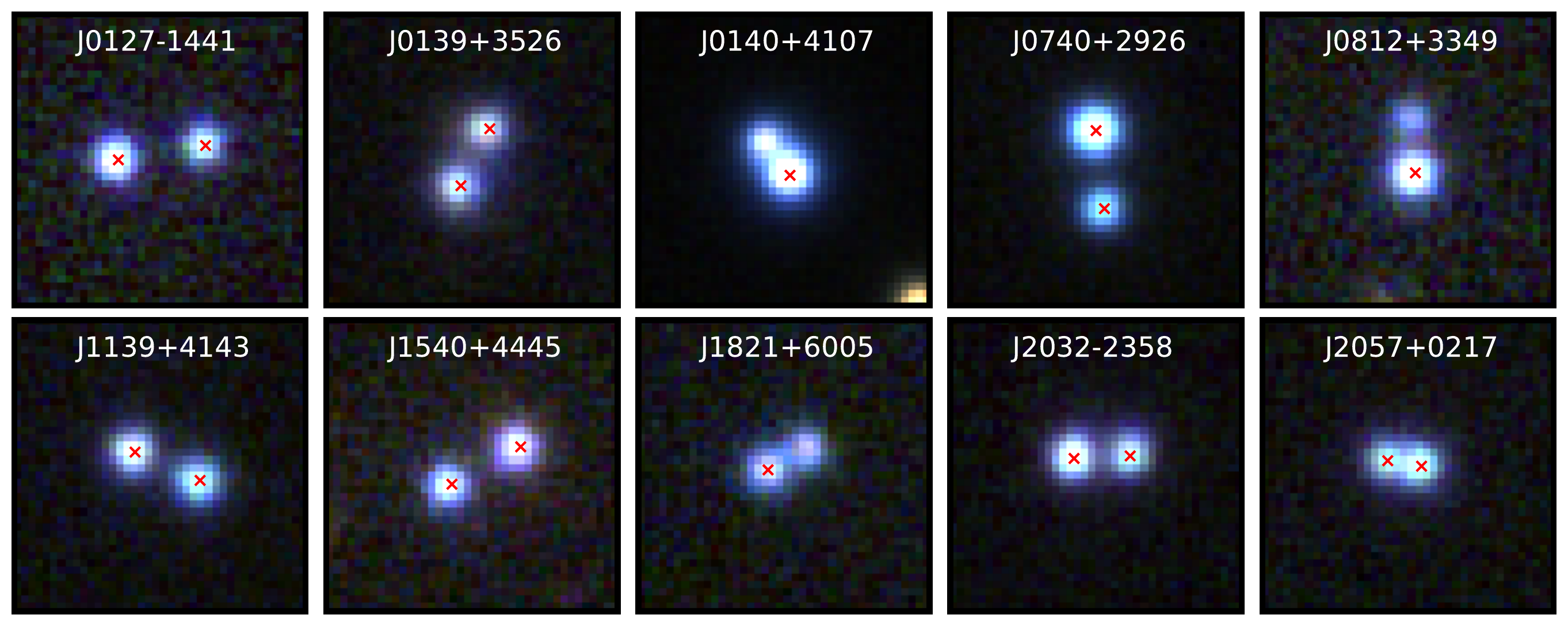}
    \caption{Pan-STARRS \textit{gri} colour images of the nearly identical quasars. Cutouts are 10$\arcsec$.}
    \label{fig:overlaid_gri_binary}
\end{figure}

\begin{center}
\begin{figure*}
	\includegraphics[width=\textwidth]{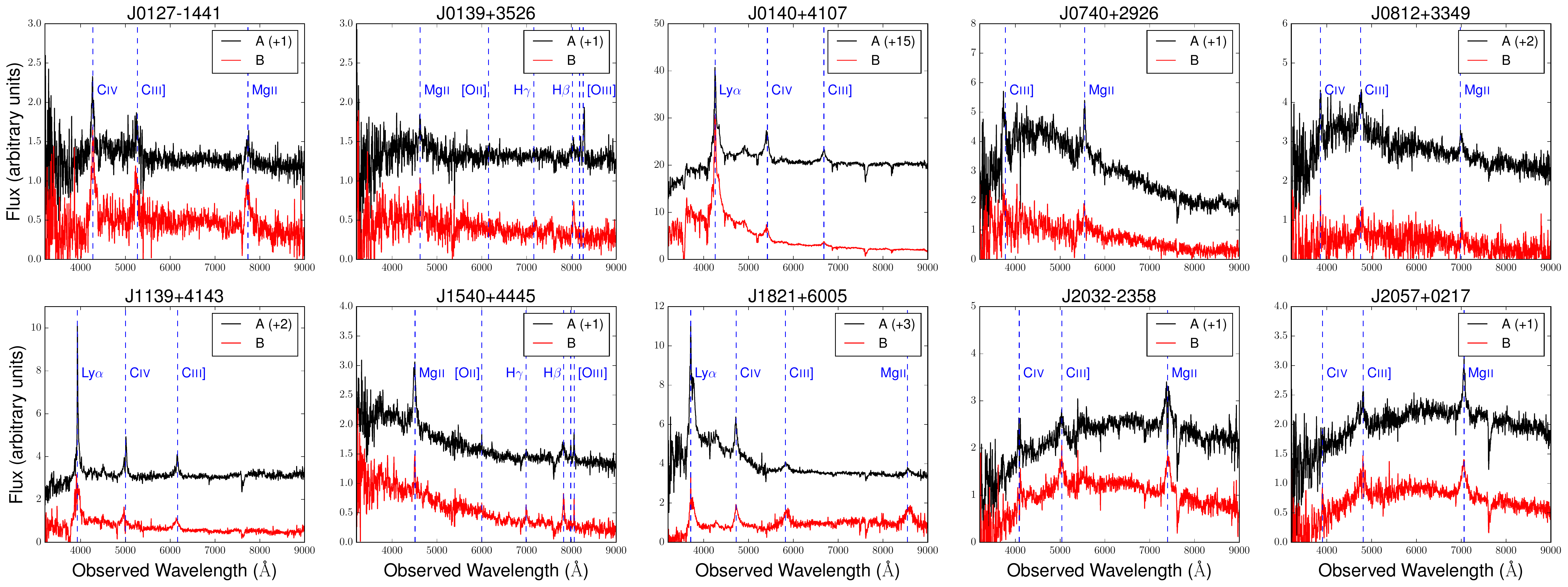}
    \caption{Spectra of the nearly identical quasars and binaries.}
    \label{fig:binary_spectra}
\end{figure*}
\end{center}

We further discover 10 systems consisting of pairs of quasars at the same redshift, shown in Figure \ref{fig:overlaid_gri_binary}. Their spectra are shown in Figure \ref{fig:binary_spectra}. However in these systems, the residuals after PSF subtraction do not convincingly demonstrate a lens, or the spectral features rule out the lensing hypothesis. See section \ref{notes} for details on individual systems.

\begin{table*}
	\centering
	\caption{Summary of observations. NIQ=nearly identical quasar, assigned to systems of quasars at the same redshift but without photometric detection of a lensing galaxy. Selection shows the quasar catalogue and \Gaia technique with which the candidate was selected: S=single, D=double, T=triple. All dates are in 2017. WISE magnitudes and colours are given in the AB system.}
	\label{tab:observations}
	\begin{tabular}{lcccccll}
		\hline
		Name & RA & DEC & Selection & $W1-W2$, $W1$& \Gaia G & Date, Exp. Time & Outcome\\
		\hline
		J0003+4555 & 0.96401 & 45.92215 & D+WISE & 0.37, 17.95 & 17.61, 18.97 & 12 Sep, 600s & quasar+star \\
		J0011\MINUS 0845 & 2.83435 & -8.76407 & D+GMM/WISE & 0.10, 17.89 & 20.31, 20.35 & 12 Sep, 1200s & \textbf{lens}, z=1.70 \\
        J0028+0631 & 7.09369 & 6.53195 & S+GMM/WISE & 0.16, 16.86 & 18.95 & 12 Sep, 1200s & \textbf{lens}, z=1.06 \\
		J0030\MINUS 1525 & 7.56378 & -15.41752 & S+WISE & -0.09, 17.01 & 19.30 & 13 Sep, 1200s & \textbf{quad lens}, z=3.36 \\
		J0123\MINUS 0455 & 20.84084 & -4.93266 & S+GMM/WISE & 0.21, 17.28 & 20.29 & 12 Sep, 2100s & \textbf{lens}, z=1.38 \\
		J0127\MINUS 1441 & 21.78539 & -14.68861 & D+WISE & 0.61, 18.14 & 20.15, 20.50 & 12 Sep, 1200s & NIQ, z=1.76 \\
		J0139+3526 & 24.88888 & 35.43658 & D+WISE & 0.47, 16.52 & 19.46, 19.65 & 12 Sep, 600s & NIQ, z=0.65 \\
        J0140\MINUS 1152 & 25.01231 & -11.872 & S+WISE & 0.19, 17.07 & 18.41 & 12 Sep, 1800s & \textbf{lens}, z=1.80 \\
        J0140+4107 & 25.20420 & 41.13331 & S+WISE & 0.27, 16.82 & 17.54 & 12 Sep, 1200s & NIQ, z=2.50 \\
        J0146\MINUS 1133 & 26.63691 & -11.56113 & D+WISE & 0.46, 16.87 & 18.39, 18.66 & 12 Sep, 1800s & \textbf{lens}, z=1.44 \\
        J0232\MINUS 2429 & 38.06565 & -24.49433 & S+WISE & 0.48, 17.14 & 17.99 & 13 Sep, 600s & quasar+star \\
        J0235\MINUS 2433 & 38.86431 & -24.55356 & D+WISE & 0.34, 16.70 & 18.12, 18.85 & 12 Sep, 1800s & \textbf{lens}, z=1.44 \\
        J0259\MINUS 2338 & 44.88961 & -23.63388 & D+WISE & 0.13, 16.91 & 19.23, 20.34 & 12 Sep, 2400s & \textbf{lens}, z=1.19 \\
        J0322+5024 & 50.71298 & 50.41402 & D+WISE & 0.17, 17.49 & 18.82, 19.11 & 12 Sep, 600s & stars \\
        J0417+3325 & 64.49682 & 33.41700 & S+WISE & 0.34, 16.87 & 19.22 & 13 Sep, 1200s & \textbf{lens}, z=1.41 \\
        J0511\MINUS 0351 & 77.91098 & -3.85049 & D+WISE & 0.57, 18.20 & 19.25, 19.99 & 13 Sep, 600s & quasar+other \\
        J0515+0652 & 78.75772 & 6.86855 & S+WISE & 0.35, 18.15 & 18.98 & 13 Sep, 1200s & quasar+other \\
        J0616+4912 & 94.13678 & 49.20712 & S+WISE & 0.45, 16.49 & 18.33 & 12 Sep, 600s & star+quasar \\
        J0630\MINUS 1201 & 97.53796 & -12.02223 & T+GMM & 0.24, 16.89 & 19.61, 19.76, 19.95 & 01 Apr, 1200s & \textbf{5-image lens}, $z=3.34$\\
        J0659+5217 & 104.92159 & 52.28907 & D+WISE & 0.13, 17.99 & 18.60, 19.63 & 12 Sep, 600s & stars \\
        J0723+4739 & 110.93660 & 47.65259 & S+WISE & 0.39, 17.35 & 19.60 & 12 Sep, 1200s & inconclusive \\
        J0740+2926 & 115.05603 & 29.44677 & D+GMM/WISE & 0.66, 16.81 & 18.20, 19.64 & 12 Sep, 1200s & NIQ, z=0.98 \\
        J0812+3349 & 123.22844 & 33.83062 & S+GMM & 0.55, 18.22 & 19.75 & 13 Sep, 1500s & NIQ, z=1.49 \\
        J0822+6659	& 125.57509 & 66.99985 & D+GMM/WISE & 0.29, 16.75 & 18.77, 19.20 & 31 Mar, 900s & stars \\
        J0823+4929	& 125.87600 & 49.48748 & D+GMM/WISE & 0.17, 16.99 & 19.41, 19.99 & 01 Apr, 600s & quasar pair, $z=0.52, 0.86$ \\
        J0826+7002 & 126.53489 & 70.04488 & S+WISE & 0.03, 15.90 & 17.50 & 13 Sep, 600s & inconclusive, quasar(+star?) \\
        J0840+3550	& 130.13842 & 35.83334 & S+GMM/WISE & -0.12, 17.21 & 19.95 & 31 Mar, 1200s & \textbf{lens}, $z=1.77$, $z_{lens}$=0.26 \\
        J0941+0518	& 145.34378 & 5.30664 & SDSS spectra & -0.05, 16.68 & --- & 31 Mar, 1200s & \textbf{lens}, $z=1.54$, $z_{lens}$=0.34 \\
        J0949+4208	& 147.47830 & 42.13381 & D+GMM/WISE & 0.14, 16.72 & 18.94, 19.81 & 31 Mar, 1200s & \textbf{lens}, $z=1.27$, $z_{lens}=0.51$ \\
        J1139+4143	& 174.94610 & 41.73088 & D+GMM/WISE & 0.32, 18.05 & 19.46, 19.62 & 01 Apr, 1200s & NIQ, $z=2.23$ \\
        J1147+3634	& 176.89300 & 36.57819 & D+GMM & 0.79, 18.62 & 19.16, 20.23 & 31 Mar, 600s & quasar + star \\
        J1239\MINUS 2216 & 189.83645 & -22.27778 & S+GMM/WISE & 0.61, 17.81 & 18.28 & 01 Apr, 750s & quasar + galaxy \\
        J1440+3736	& 220.20396 & 37.61107 & S+WISE & 0.43, 16.91 & 19.27 & 01 Apr, 600s & galaxy + star \\
        J1508+3844	& 227.18253 & 38.73934 & D+GMM/WISE & 0.11, 17.28 & 20.21, 20.92 & 31 Mar, 2700s & \textbf{lens}, $z=1.68$ \\
        J1536+3629	& 234.01479 & 36.49226 & S+GMM & -0.36, 19.93 & 20.51 & 01 Apr, 600s & galaxy at $z=0.111$ \\
        J1540+4445	& 235.10759 & 44.75457 & D+GMM/WISE & 0.09, 17.37 & 19.73, 20.37 & 01 Apr, 600s & NIQ,  $z=0.61$\\
        J1551+3157	& 237.77584 & 31.95027 & S+GMM/WISE & 0.04, 17.41 & 20.48 & 31 Mar, 1200s & inconclusive, ($z=2.27$?) \\
        J1554+2616	& 238.54871 & 26.27657 & D+GMM/WISE & 0.07, 17.88 & 18.95, 20.03 & 31 Mar, 600s & quasar + star  \\
        J1602+4526	& 240.70535 & 45.43528 & S+GMM & -0.23, 17.50 & 20.17 & 31 Mar, 2700s & \textbf{lens}, $z=2.16$, $z_{lens}=0.43$ \\
        J1606\MINUS 2333 & 241.50074 & -23.55612 & D+WISE & 0.64, 16.42 & 18.74, 18.88 & 31 Mar, 2400s & \textbf{quad lens}, $z=1.69$ \\
        J1611+5756	& 242.98266 & 57.93872 & S+GMM/WISE & 0.09, 17.74 & 20.24 & 31 Mar, 600s & galaxies at $z=0.257$ \\
        J1617\MINUS 2146 & 244.25462 & -21.76683 & D+WISE & 0.65, 18.15 & 19.75, 20.39 & 01 Apr, 1500s & inconclusive, likely stars \\
        J1617\MINUS 2305 & 244.34009 & -23.09620 & D+WISE & 0.73, 18.19 & 18.95, 19.23 & 13 Sep, 1200s & quasar+star \\
        J1640+1045	& 250.07549 & 10.75175 & D+GMM/WISE & 0.13, 17.02 & 18.12, 19.87 & 31 Mar, 3000s & \textbf{lens}, $z=1.7$ \\
        J1709+3828	& 257.36966 & 38.46700 & D+WISE & -0.01, 17.31 & 20.07, 20.33 & 31 Mar, 2650s & \textbf{lens}, $z=1.38$ \\
        J1710+4332	& 257.74257	 & 43.54287 & S+GMM & -0.50, 18.27 & 20.49 & 31 Mar, 2100s & \textbf{lens}, $z=3.08$ \\
		J1721+8842 & 260.45419 & 88.70621 & D+WISE & 0.28, 15.65 & 17.97, 18.24 & 13 Sep, 4800s & \textbf{quad lens}, $z=2.37$ \\
		J1821+6005 & 275.37642 & 60.09062 & S+WISE & 0.53, 18.20 & 19.96 & 13 Sep, 1800s & NIQ, z=2.05 \\
		J1831+5447 & 277.86360 & 54.79965 & D+WISE & 0.10, 16.23 & 18.79, 19.86 & 12 Sep, 1200s & \textbf{lens}, z=1.07 \\
		J2018\MINUS 3015 & 304.73256 & -30.26574 & T+WISE & 0.32, 16.21 & 18.26, 18.52, 19.72 & 12 Sep, 900s & stars+galaxy \\
		J2032\MINUS 2358 & 308.15741 & -23.97291 & D+WISE & 0.72, 17.59 & 19.04, 19.17 & 13 Sep, 1200s & NIQ, z=1.64 \\
		J2057+0217 & 314.46696 & 2.29683 & D+GMM/WISE & 0.81, 18.17 & 20.06, 20.10 & 12 Sep, 1200s & NIQ, z=1.52 \\
		J2058\MINUS 0744 & 314.53051 & -7.74705 & D+WISE & 0.51, 17.97 & 19.72, 19.75 & 13 Sep, 600s & quasar+star \\
		J2111+1349 & 317.80707 & 13.82978 & S+WISE & 0.70, 17.83 & 19.74 & 13 Sep, 600s & quasar+star \\
		J2124+1632 & 321.07029 & 16.53841 & S+GMM/WISE & 0.40, 16.48 & 19.11 & 12 Sep, 1200s & \textbf{lens}, z=1.28  \\
		J2302\MINUS 2813 & 345.74028 & -28.22055 & S+WISE & 0.80, 17.67 & 18.40 & 12 Sep, 600s & quasar+star \\
		J2305+3714 & 346.48273 & 37.23932 & D+WISE & 0.25, 16.35 & 17.55, 18.71 & 12 Sep, 1200s & \textbf{lens}, z=1.78 \\
		J2327+2238 & 351.75343 & 22.63698 & D+GMM/WISE & 0.12, 17.38 & 20.68, 20.75 & 13 Sep, 600s & quasar pair, z=0.53, 0.55 \\
		J2332\MINUS 1852 & 353.08034 & -18.86853 & S+WISE & 0.32, 17.44 & 19.48 & 12 Sep, 1200s & \textbf{lens}, z=1.49 \\
		J2350\MINUS 1930 & 357.58645 & -19.51585 & D+WISE & 0.50, 18.08 & 19.49, 20.71 & 13 Sep, 600s & quasar+star \\
		\hline
	\end{tabular}
\end{table*}

\section{Modelling} \label{modelling}
\subsection{Pixel modelling} \label{pixelmodelling}
In this section we model the Pan-STARRS \textit{grizY} imaging data for each lens system to derive simple component shapes, positions and colours. In the next section we use the image and galaxy positions and flux measurements to fit simple lens models to each system.

The PSF is derived by fitting a Moffat profile \citep{moffat1969} to a nearby star. The \textit{grizY} Pan-STARRS images are modelled simultaneously with each quasar image fit with a PSF, and galaxies fit with S\'ersic profiles \citep{sersic1963} convolved with the PSF. In all lens systems, the presence of a lens galaxy is apparent from the colour image, except for J0630-1201 and J1606-2333. The free parameters for the pixel modelling are the positions of the quasar images and the lensing galaxy, as well as the flattening, size and S\'ersic index of the galaxy, all of which we assume to be the same across bands. The log likelihood is sampled using the \textit{emcee} package \citet{foreman-mackey2013} to determine statistical uncertainties. Finally the fitting is repeated with a different PSF star to determine the systematic error for a possible PSF mismatch.

\begin{figure*}
\centering
    \caption{Pixel modelling of the confirmed lenses. Left to right: \textit{gri} data, model, \textit{gri} PSF-subtracted, and r-band residuals. Blue crosses indicate the positions of quasar images and red plusses mark the locations of lensing galaxies.}
    \begin{tabular}{ll}
    \includegraphics[scale=0.38]{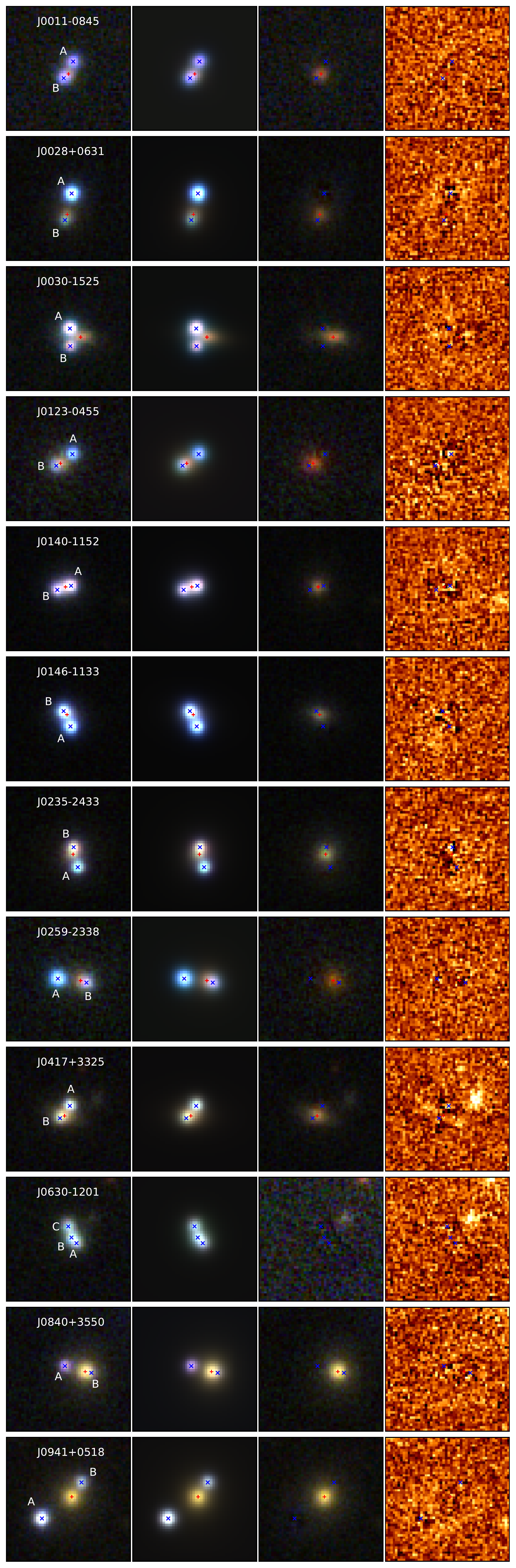}
    &
    \includegraphics[scale=0.38]{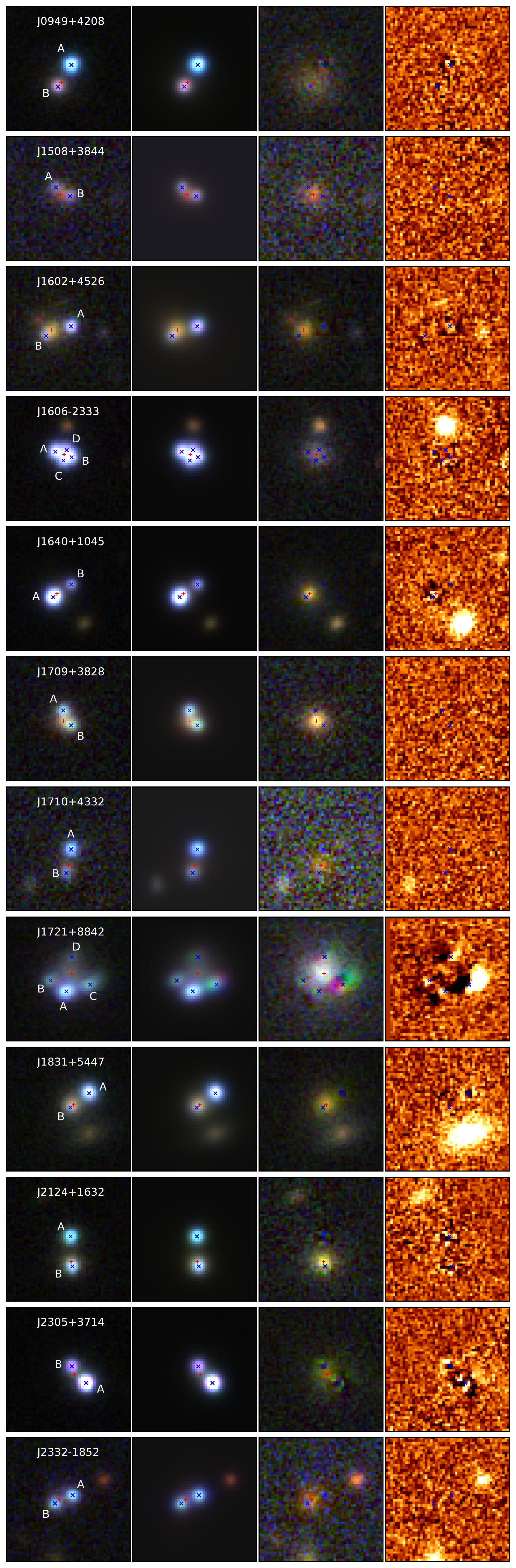}
    \end{tabular}
    \label{fig:psf_subtraction}
\end{figure*}

Figure \ref{fig:psf_subtraction} shows the Pan-STARRS \textit{gri} images and the residuals after modelling each system as the relevant sum of quasars and galaxies. Astrometry and photometry for all available bands are given in Table \ref{tab:astrophotometry}.

\subsection{Mass modelling}
We fit all the systems with singular isothermal elipsoids (SIE) using our own Python-based image-plane modelling code, which has been checked for consistency with \textit{lensmodel} \citep{keeton2001}. We use the two-step process of modelling the pixels to find positions and fluxes, and then deriving a lens model. This is to ensure that, if the system is not well-described by a singular isothermal ellipsoid, it does not affect the PSF subtraction. Furthermore we can quantify the chi-squared contributions from positions and flux ratios robustly. For all the doubly imaged lenses, we start from two image positions and a galaxy position (6 parameters). However to constrain a singular isothermal ellipsoid (galaxy position, mass, ellipticity, position angle and source position; 7 parameters), we require one further observable. For this we use the image flux ratio. We take the median flux ratio from the \textit{griz} bands and include a 20 percent uncertainty on the input fluxes since optical flux ratios depend not only on the lensing geometry but also on extinction, quasar variability over the time delay, microlensing, etc. Using these pixel-based measurements and their uncertainties, the lens parameters are inferred through image-plane sampling. Given the extra information in quads, we are able to use more realistic models with more parameters. We use SIE+shear models for these systems.

The lens model parameters (medians with 1$\sigma$ limits) and chi-squared contributions are listed in Table \ref{tab:massmodels}. For the three quadruply imaged lensed quasars we include a 50 percent uncertainty on the input fluxes for the two saddle point images in each system, in accordance with their increased susceptibility to microlensing \citep{schechter2002}.

Systems with significantly elliptical SIE fits and a large image-position chi-squared are indicative of a strong external shear if the lens is not elliptical in the photometry. 

\begin{table*}
	\centering
	\caption{Median parameter values with 1$\sigma$ uncertainties for mass models and light profiles. b=Einstein radius, PA= position angle (north of west), q = axis ratio, $\mu$ = total source magnification. $^{\dagger}$ The mass model and galaxy light profile for J0030-1525 are based on better seeing VST-ATLAS data \citep{shanks2015}  as described in Section \ref{notes}.}
	\label{tab:massmodels}
	\begin{tabular}{cccccccccc}
		\hline
		name & $b$ ($\arcsec$) & $PA_{SIE}$ &$q_{SIE}$&  $PA_{light}$ &$q_{light}$& ${{\chi}^2}_{gal}$ & ${{\chi}^2}_{images}$ & ${{\chi}^2}_{flux}$ & $\mu$  \\
		\hline
        \noalign{\vskip 0.5mm}
        J0011\MINUS 0845 & $0.96_{0.95}^{0.97}$ & $176_{174}^{177}$ & $0.70_{0.68}^{0.73}$ & $99_{55}^{131}$ & $0.86_{0.73}^{0.97}$ & 0.07 & 0.19 & 0.03 & $5.0_{4.5}^{5.3}$ \\
        \noalign{\vskip 0.5mm}
        J0028+0631 & $1.43_{1.42}^{1.44}$ & $55_{51}^{57}$ & $0.81_{0.79}^{0.83}$ & $58_{54}^{62}$ & $0.86_{0.84}^{0.88}$ & 0.07 & 0.19 & 0.02 & $4.2_{4.1}^{4.4}$ \\
        \noalign{\vskip 0.5mm}
        \ J0030\MINUS 1525$^{\dagger}$ & $1.08_{1.05}^{1.15}$ & $170_{165}^{182}$ & $0.82_{0.33}^{0.95}$ & $55_{51}^{57}$ & $0.81_{0.79}^{0.83}$ & 0.97 & 13.0 & 13.1 & $71_{29}^{88}$ \\
        \noalign{\vskip 0.5mm}
        J0123\MINUS 0455 & $0.96_{0.95}^{0.97}$ & $136_{134}^{140}$ & $0.72_{0.68}^{0.75}$ & $6_{1}^{10}$ & $0.84_{0.82}^{0.86}$ & 0.04 & 0.09 & 0.01 & $3.5_{3.4}^{3.7}$ \\
        \noalign{\vskip 0.5mm}
        J0140\MINUS 1152 & $0.72_{0.71}^{0.73}$ & $127_{120}^{149}$ & $0.56_{0.54}^{0.57}$ & $101_{94}^{113}$ & $0.92_{0.89}^{0.95}$ & 0.07 & 0.18 & 0.01 & $10.3_{9.3}^{11.4}$ \\
        \noalign{\vskip 0.5mm}
        J0146\MINUS 1133 & $0.83_{0.82}^{0.84}$ & $4.5_{3.4}^{5.4}$ & $0.52_{0.45}^{0.58}$ & $175_{174}^{176}$ & $0.43_{0.41}^{0.45}$ & 0.03 & 0.11 & 0.01 & $3.9_{3.8}^{4.0}$ \\
        \noalign{\vskip 0.5mm}
        J0235\MINUS 2433 & $1.04_{1.03}^{1.05}$ & $44_{41}^{45}$ & $0.74_{0.73}^{0.75}$ & $68_{65}^{70}$ & $0.89_{0.88}^{0.91}$ & 0.06 & 0.19 & 0.02 & $5.4_{5.3}^{5.5}$ \\
        \noalign{\vskip 0.5mm}
        J0259\MINUS 2338 & $1.41_{1.40}^{1.42}$ & $99_{97}^{101}$ & $0.67_{0.65}^{0.70}$ & $86_{28}^{148}$ & $0.98_{0.96}^{1.00}$ & 0.06 & 0.17 & 0.04 & $3.1_{3.0}^{3.2}$ \\
        \noalign{\vskip 0.5mm}
        J0417+3325 & $0.82_{0.81}^{0.83}$ & $173_{171}^{175}$ & $0.65_{0.64}^{0.68}$ & $177_{176}^{178}$ & $0.48_{0.46}^{0.49}$ & 0.06 & 0.19 & 0.01 & $4.0_{3.9}^{4.1}$ \\
        \noalign{\vskip 0.5mm}
        J0840+3550 & $1.43_{1.39}^{1.45}$ & $164_{160}^{166}$ & $0.83_{0.79}^{0.88}$ & $89_{85}^{93}$ & $0.88_{0.87}^{0.90}$ & 0.12 & 0.72 & 0.39 & $4.5_{4.1}^{4.9}$ \\
        \noalign{\vskip 0.5mm}
        J0941+0518 & $2.72_{2.67}^{2.73}$ & $97_{93}^{109}$ & $0.81_{0.77}^{0.82}$ & $66_{64}^{68}$ & $0.88_{0.87}^{0.89}$ & 0.02 & 0.12 & 0.01 & $5.2_{5.1}^{5.3}$ \\
        \noalign{\vskip 0.5mm}
        J0949+4208 & $1.23_{1.21}^{1.30}$ & $155_{152}^{157}$ & $0.68_{0.62}^{0.76}$ & $137_{119}^{149}$ & $0.96_{0.94}^{0.98}$ & 0.08 & 0.12 & 0.02 & $3.1_{2.9}^{3.2}$ \\
        \noalign{\vskip 0.5mm}
        J1508+3844 & $0.92_{0.90}^{0.93}$ & $103_{100}^{107}$ & $0.91_{0.88}^{0.93}$ & $173_{144}^{192}$ & $0.88_{0.79}^{0.96}$ & 0.10 & 0.25 & 0.03 & $19_{15}^{25}$ \\
        \noalign{\vskip 0.5mm}
        J1602+4526 & $1.41_{1.40}^{1.44}$ & $78_{75}^{80}$ & $0.59_{0.57}^{0.61}$ & $143_{140}^{145}$ & $0.74_{0.72}^{0.75}$ & 0.12 & 0.19 & 0.05 & $3.4_{3.3}^{3.5}$ \\
        \noalign{\vskip 0.5mm}
        J1606\MINUS 2333 & $0.60_{0.59}^{0.61}$ & $95_{87}^{115}$ & $0.77_{0.69}^{0.82}$ & $130_{104}^{173}$ & $0.67_{0.51}^{0.95}$ & 0.21 & 9.9 & 0.54 & $9.4_{8.5}^{10.7}$ \\
        \noalign{\vskip 0.5mm}
        J1640+1045 & $1.04_{1.03}^{1.05}$ & $122_{118}^{123}$ & $0.31_{0.30}^{0.35}$ & $119_{114}^{124}$ & $0.88_{0.86}^{0.90}$ & 0.05 & 0.07 & 0.01 & $7.6_{5.6}^{8.2}$ \\
        \noalign{\vskip 0.5mm}
        J1709+3828 & $0.96_{0.95}^{0.97}$ & $170_{167}^{171}$ & $0.81_{0.79}^{0.83}$ & $175_{170}^{181}$ & $0.81_{0.78}^{0.84}$ & 0.05 & 0.06 & 0.01 & $10.7_{9.8}^{11.9}$ \\
        \noalign{\vskip 0.5mm}
        J1710+4332 & $1.21_{1.20}^{1.23}$ & $62_{23}^{129}$ & $0.95_{0.93}^{0.98}$ & $17_{-15}^{40}$ & $0.85_{0.72}^{0.96}$ & 0.03 & 0.07 & 0.02 & $5.2_{4.6}^{6.1}$ \\
        \noalign{\vskip 0.5mm}
        J1721+8842 & $1.99_{1.98}^{2.00}$ & $169_{146}^{173}$ & $0.87_{0.78}^{0.96}$ & $136_{133}^{139}$ & $0.87_{0.86}^{0.88}$ & 2.55 & 0.64 & 1.2 & $27_{23}^{33}$ \\
        \noalign{\vskip 0.5mm}
        J1831+5447 & $1.10_{1.09}^{1.11}$ & $126_{122}^{130}$ & $0.75_{0.71}^{0.79}$ & $44_{41}^{48}$ & $0.87_{0.85}^{0.90}$ & 0.06 & 0.18 & 0.02 & $3.1_{3.0}^{3.2}$ \\
        \noalign{\vskip 0.5mm}
        J2124+1632 & $1.41_{1.40}^{1.42}$ & $176_{173}^{177}$ & $0.39_{0.34}^{0.50}$ & $148_{145}^{153}$ & $0.75_{0.73}^{0.78}$ & 0.07 & 0.11 & 0.01 & $2.7_{2.6}^{2.8}$ \\
        \noalign{\vskip 0.5mm}
        J2305+3714 & $1.10_{1.09}^{1.11}$ & $59_{56}^{62}$ & $0.62_{0.60}^{0.67}$ & $91_{85}^{98}$ & $0.88_{0.85}^{0.90}$ & 0.05 & 0.07 & 0.01 & $8.6_{8.2}^{9.6}$ \\
        \noalign{\vskip 0.5mm}
        J2332\MINUS 1852 & $0.96_{0.95}^{0.97}$ & $99_{96}^{101}$ & $0.63_{0.59}^{0.67}$ & $75_{72}^{78}$ & $0.51_{0.47}^{0.55}$ & 0.05 & 0.13 & 0.01 & $4.5_{4.3}^{4.9}$ \\
		\hline
	\end{tabular}
\end{table*}

\section{Discussion} \label{discussion}
\subsection{Notes on individual systems} \label{notes}
\begin{table}
	\centering
	\caption{Summary of lensed quasars (image separations are the largest ones for quads). $^{\dagger}$ Lens redshift from \citet{agnellovst2017}. *These lenses have been independently identified as candidates or confirmed lenses by others as described in Section \ref{notes}.}
	\label{tab:lenssummary}
	\begin{tabular}{ccccc}
		\hline
		name & $z_{source}$ & $z_{lens}$ & sep. (\arcsec) & $i_{images}$, $i_{lens}$ \\
		\hline
        J0011\MINUS 0845 & 1.70 & - & 1.92 & 20.09, 20.32 \\
        J0028+0631 & 1.06 & - & 2.78 & 18.81, 18.44 \\
        J0030\MINUS 1525 & 3.36 & - & 1.78 & 18.60, 18.88 \\
        J0123\MINUS 0455 & 1.38 & - & 2.00 & 19.42, 18.15 \\
        \ J0140\MINUS 1152* & 1.80 & 0.28$^{\dagger}$ & 1.45 & 18.25, 18.53 \\
        \ J0146\MINUS 1133* & 1.44 & - & 1.69 & 18.09, 18.70 \\
        \ J0235\MINUS 2433* & 1.44 & - & 2.05 & 18.74, 18.10 \\
        \ J0259\MINUS 2338* & 1.19 & - & 2.92 & 19.17, 18.72 \\
        J0417+3325 & 1.41 & - & 1.59 & 18.89, 18.54 \\
        J0630\MINUS 1201 & 3.34 & - & 1.90 & 18.30,\ \ \  --- \ \  \\
        J0840+3550 & 1.77 & 0.26 & 2.46 & 19.74, 18.03 \\
        \ J0941+0518* & 1.54 & 0.34 & 5.40 & 18.30, 17.51 \\
        J0949+4208 & 1.27 & 0.51 & 2.57 & 18.88, 19.19 \\
        J1508+3844 & 1.68 & - & 1.69 & 20.67, 20.04 \\
        J1602+4526 & 2.16 & 0.43 & 2.70 & 19.81, 18.67 \\
        J1606\MINUS 2333 & 1.69 & - & 1.74 & 17.58, 20.85 \\
        \ J1640+1045* & 1.70 & - & 2.22 & 18.24, 18.67 \\
        J1709+3828 & 1.38 & - & 1.70 & 19.90, 19.36 \\
        J1710+4332 & 3.08 & - & 2.43 & 20.95, 20.60 \\
        \ J1721+8842* & 2.37 & - & 4.03 & 18.36, 18.02\\
        J1831+5447 & 1.07 & - & 2.32 & 18.80, 18.17 \\
        J2124+1632 & 1.28 & - & 3.02 & 18.11, 18.40 \\
        J2305+3714 & 1.78 & - & 2.20 & 17.03, 18.32\\
        J2332\MINUS 1852 & 1.49 & - & 1.97 & 18.76, 18.93 \\
		\hline
	\end{tabular}
\end{table}

\begin{table}
	\centering
	\caption{Summary of nearly identical quasars and binaries.}
	\label{tab:NIQsummary}
	\begin{tabular}{cccc}
		\hline
		name & $z_{source}$ & sep. (\arcsec) & $i_{mag}$ \\
		\hline
        J0127\MINUS 1441 & 1.76 &  2.96 & 19.30 \\
        J0139+3526 & 0.65 & 2.22 & 18.49 \\
        J0140+4107 & 2.50 & 1.44 & 16.94 \\
        J0740+2926 & 0.98 & 2.59 & 18.23 \\
        J0812+3349 & 1.49 & 1.99 & 19.12 \\
        J1139+4143 & 2.23 & 2.30 & 18.86 \\
        J1540+4445 & 0.61 & 2.74 & 19.21 \\
        J1821+6005 & 2.05 & 1.48 & 19.34 \\
        J2032\MINUS 2358 & 1.64 & 1.91 & 18.40 \\
        J2057+0217 & 1.52 & 1.06 & 18.96 \\
		\hline
	\end{tabular}
\end{table}

\subsubsection{J0030\MINUS 1525}
\begin{center}
\begin{figure}
\hspace*{-0.3cm}
	\includegraphics[width=\columnwidth]{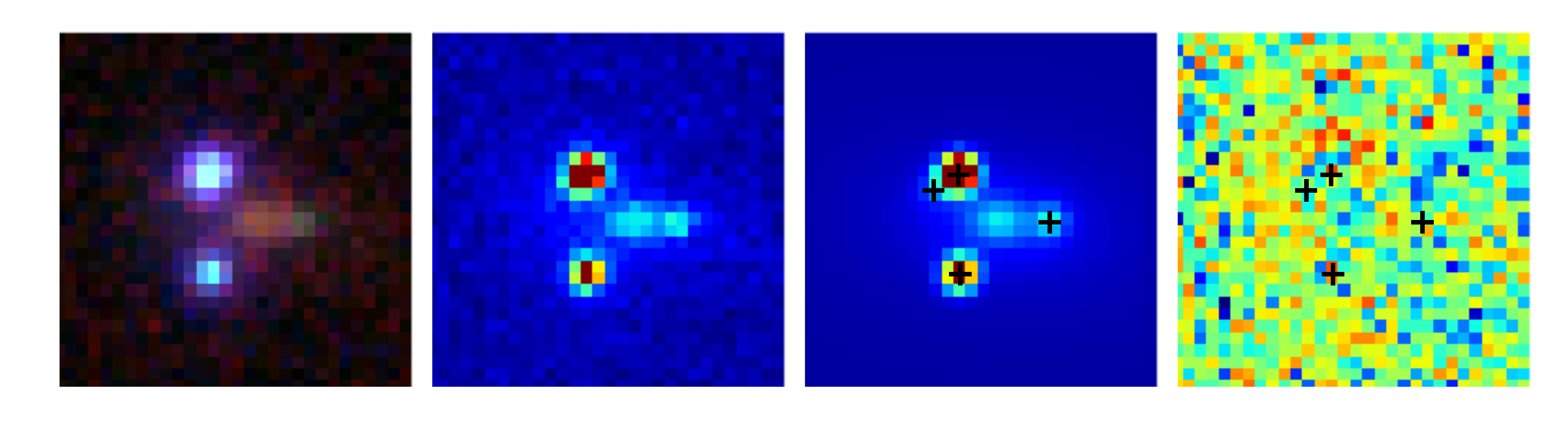}
    \caption{From the left: PSJ0030-1525 VST \textit{grz} colour image; 0.42\arcsec -seeing r-band VST image; 4PSFs+galaxy model with PSF positions overlaid; residuals after model subtraction.}
    \label{fig:J0030_VST}
\end{figure}
\end{center}

The Pan-STARRS \textit{gri} image for this object shows two bright PSFs and a galaxy significantly offset from where it should lie to create a double-image system. However a better seeing VST r-band image, Figure \ref{fig:J0030_VST}, resolves four objects including a faint blue PSF next to an extended red object. This is consistent with a flux-ratio anomaly fold-configuration quad with the faint PSF being the counterimage. We model the Pan-STARRS data for photometry of A+B, C and D+G, given in Table \ref{tab:astrophotometry}. However for the mass model we use the VST r-band data given its excellent seeing and resolution of the counterimage. The PSF is inferred from the data due to the lack of a nearby star. We find that the data are fit by models with the merging pair consisting of a bright image north-west of a faint image or vice versa. The former is much more plausible given that saddle points are more commonly demagnified than maxima and minima \hbox{\citep{schechter2002}} and so this is the mass model we report in Table \ref{tab:massmodels}. The best fit mass model gives flux ratios of $\sim$ 7:7:3:1 while the measured flux ratios are $\sim$ 7:0.5:4:1, i.e., a 14$\times$ decrease for image B.

\subsubsection{J0127\MINUS 1441}
DECaLS data reveal two faint red objects between the two PSFs that could be responsible for the lensing. These are only detected in the $z$-band data and each has a magnitude of $\sim$ 24. Comparing to mock lenses from \citet{oguri2010} with similar source redshift and image separation, the faintest $i$-band lens magnitudes are $\sim$ 22. While the multiple component nature of this putative lens complicates the comparison, the faintness and large separation suggests either a binary quasar or high-redshift group lens. Deeper data will help secure the magnitudes of these objects.

\subsubsection{J0139+3526}
While this was a promising candidate given the similar SEDs and residuals consistent with a lensing galaxy, the redshift of the quasars is low ($z=0.65$) and the [O\textsc{iii}] emission lines are significantly different. There is no discernible redshift difference from the spectra, so this system is likely a quasar merger in which we are seeing the onset of AGN activity. The residuals are probably associated with the quasar host galaxies interacting. A nearby bright star makes this system ideal for adaptive optics follow-up.

\subsubsection{J0140-1152}
This lens has been independently identified in the VST-ATLAS survey by \citet{agnellovst2017}, who report a lens redshift of 0.277. It is a highly magnified ($\sim$ 10 times) double. The elliptical mass model suggests a strong external shear given the relatively circular light profile of the galaxy.

\subsubsection{J0235-2433}
\citet{agnello2017des} have independently selected this object as a candidate gravitational lens from the Dark Energy Survey (DES) footprint \citep{despaper}. We confirm the PSFs to be quasars at a redshift of 1.44. The DES data displayed in \citet{agnello2017des} show that the image closest to the lens galaxy is brighter than the other, while the opposite is true in the Pan-STARRS data. This change by approximately one magnitude within less than a few years (mean epoch for the Pan-STARRS detection is 56475, and $\sim$57350 for DES \citep{desdatarelease}) is likely attributed to a microlensing event rather than quasar variability. Indeed in the \Gaia data, the closest and furthest images have magnitudes of G=18.12 and G=18.85 respectively. Since both images are detected by \textit{Gaia}, they will have well-sampled lightcurves over 5 years.  When released, these lightcurves will clearly distinguish a microlensing event in one image from quasar variability which would be seen in both images but separated by the time delay.

\subsubsection{J0417+3325}
  This double has an elliptical lens galaxy ($q=0.48$) and the SIE lens model is aligned along the same position angle but less elliptical ($q=0.65$). It was originally detected by \citet{colla1970} at 408MHz with a flux of 290 mJy, and subsequently at 5GHz by \citet{davis1971} and \citet{altschuler1986} with detections of 60 mJy and 39 mJy respectively. It is also detected in NVSS \citep{condon1998} at 1.4GHz with a flux density of 109mJy. Archival 8GHz VLA data show two point sources in the same positions as the optical quasar point sources with a hint of a third source south of image A. There is an extended source 3\arcsec \ west of the system, lacking any optical detection in Pan-STARRS. The A to B flux ratio at 8GHz is 1.2, while the $g$-band ratio---the magnitudes least affected by the presence of the lens galaxy---is 2.0, a discrepancy that could be explained by a number of effects including microlensing or extinction. The presence of a radio source is promising for using resolved source structure and radio flux ratios to precisely constrain lens models.

\subsubsection{J0630-1201}
This system was identified in a novel Pan-STARRS+WISE photometric quasar catalogue as explained in \hbox{\citet{ostrovski2017}}. Keck NIRC2 data reveal two lensing galaxies and a fifth demagnified image. See \hbox{\citet{ostrovski2017}} for a full mass model.

\subsubsection{J0740+2926}
This is an SDSS quasar that was subsequently found to be followed up as part of the SDSS quasar lens search, which also confirm this as a quasar pair at z=0.98 \hbox{\citep{inada2010}}.

\subsubsection{J0812+3349}
While we have classified this as a nearly identical quasar (NIQ), there is support for this object being a lens since a faint red object is seen upon subtracting the PSFs. Because of the lack of any imaging data sets other than Pan-STARRS we cannot confirm this detection since this faint residual is only seen in the i-band at $i \sim$ 23. The quasars' proximity could mean lens galaxy light is being fit or appearing due to poor PSF subtraction. Therefore deeper and/or higher-resolution imaging of this system is required.

\subsubsection{J0941+0518}
\begin{figure}
\centering
    \includegraphics[width=\columnwidth]{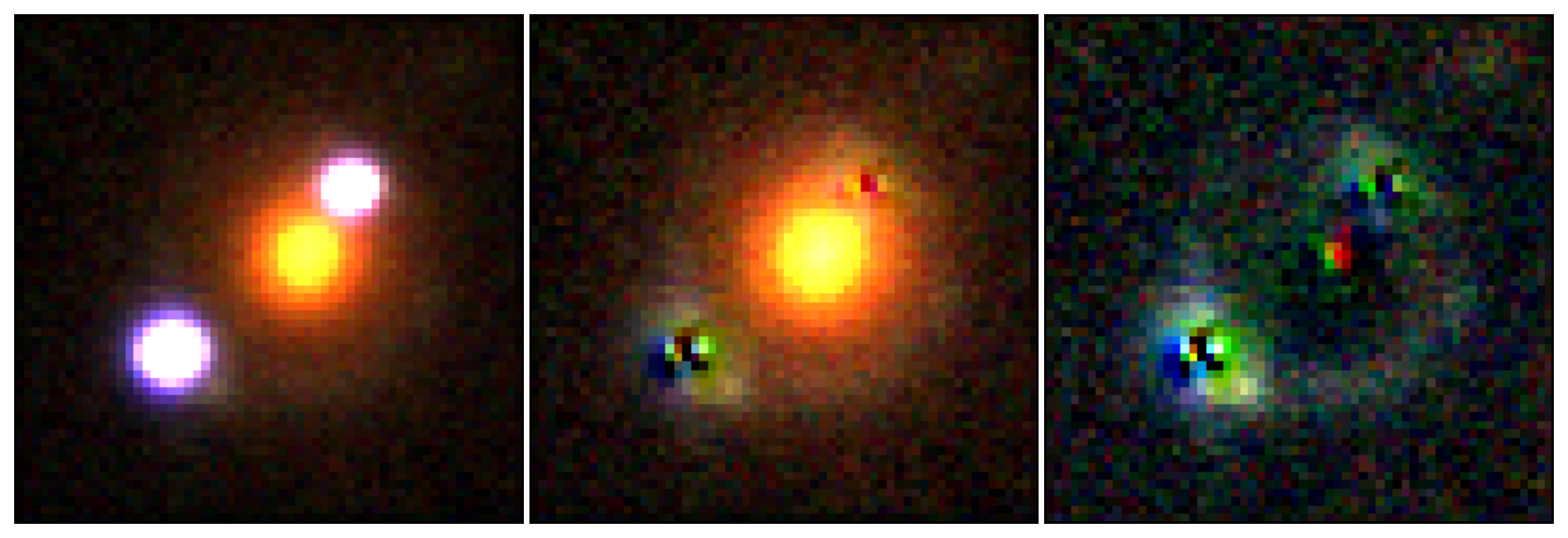}
    \caption{From left to right: J0941+0518 stacked CFHT \textit{ugr} data, PSF-subtraction, and PSF and galaxy subtraction.}
    \label{fig:J0941_cfht}
\end{figure}

This system and J1640+1045 were concurrently discovered by \citet{williams2017}.
Deep CFHT data reveals an Einstein ring of the quasar host galaxy in the $u$, $g$ and $r$ bands, as shown in Figure \ref{fig:J0941_cfht}. We calculate the velocity dispersion of our SIE model to be 365~km~s$^{-1}$ and compare this to the measured velocity dispersion of the galaxy from an SDSS spectrum of 313$\pm$18~km~s$^{-1}$. This discrepancy can be accounted for by a shallower than isothermal density profile of $\gamma \sim 1.8$ \citep[see Figure 4 of][]{auger2010}. However this lens is embedded within a galaxy group and so close companions would suggest a steeper profile \citep{dobke2007, auger2008}. Since a quasar emission line from the closer, fainter quasar image is present in the galaxy spectrum, the SDSS velocity dispersion might not be trustworthy. A deeper spectrum and deeper optical imaging of the Einstein ring will help constrain the mass model. We calculate the time delay for this system as B lagging A by $\sim$270 days.

\subsubsection{J0949+4208}
This is a radio source in NVSS and FIRST, however FIRST clearly associates it with the lensing galaxy. The lensing galaxy is likely a radio galaxy, however we cannot rule out the fainter optical image being much brighter in the radio. It was targetted for a BOSS spectrum which shows quasar emission lines at $z=1.27$ and a galaxy at $z=0.507$, in agreement with the absorption lines seen in our WHT spectrum.

\subsubsection{J1540+4445}
This NIQ is at low redshift ($z=0.61$) and shows variation in the emission line profiles between the two components. Furthermore it appears to be associated with a galaxy cluster. Two BOSS spectra of galaxies in the field place them at the same redshift as the NIQ.

\subsubsection{J1606\MINUS 2333}
This system was identified as a \Gaia double. Though the seeing for the WHT spectrum was poor, quasar emission lines are visible across the broad trace. The two-component decomposition is made the same distance as that between the \Gaia detections, and a narrow uniform aperture is used for the spectral extraction. An archival, shallow \textit{Chandra} image shows extended emission at the positions of the two \Gaia detections and also near image C, as labelled in Figure \ref{fig:psf_subtraction}. Recent HST imaging (proposal 15320, PI: Treu) clearly resolves four quasar images and a lens galaxy. The mass model suggests a shear of 0.15 and an image position chi-squared of $\sim$10. The majority of this is attributed to a poorly fitting position of image D. The best-fit models consistently place D 0.02-0.04\arcsec \ more northern than the measured value. This discrepancy could be explained by an astrometric perturbation from an unresolved galaxy near image D.

\subsubsection{J1640+1045}
This double system appears to be a flux-ratio anomaly double with the closest image 1.5 magnitudes brighter than the further image (0\farcs52 and 1\farcs71 from the galaxy respectively). While this could be explained by microlensing, the fit of an SIE is possible to explain this flux ratio but requires a highly elliptical (q $\sim$ 0.3) mass model, while the light profile is reasonably circular (q $\sim$ 0.9). Therefore if this flux ratio is to be explained by a mass model it is much more likely to be due to a strong external shear in the same direction as the inferred mass position angle. A singular isothermal sphere + shear model for this system is also a good fit (chi-squared $\sim$ 0.25), with a shear of 0.32 at 122 degrees North of East. This position angle is well-aligned with a nearby galaxy less than 4\arcsec \ from the main lensing galaxy at a position angle of 133 degrees North of East. Furthermore SDSS, Pan-STARRS, \Gaia and DECaLS data all show that this lens maintains the measured flux ratio implying this flux ratio is unchanging over 15 years. While microlensing events cannot be ruled out, it is more likely that this apparent flux-ratio anomaly is explained by a strong external shear. Since both quasar images are detected by \textit{Gaia}, their lightcurves will become useful data for breaking the microlensing/shear degeneracy for this system.

\subsubsection{J1721+8842}
The ``polar quad'' shows strong signs of line of sight absorption systems in all 4 images. The Pan-STARRS imaging data have highly distorted PSFs in some bands, so the mass model is based on flux ratios measured only in the $r$-band. The residuals after subtracting PSFs show flux to the west of image C, perhaps associated with a second lensing galaxy or an arc from the quasar host galaxy. Given its high declination, the position and airmass of this system is essentially unchanged year-round, providing an excellent opportunity to efficiently measure time delays without any seasonal gaps. We note that the mass model is well-fit to the data, with a chi-squared of $\sim$4.4 (the number of degrees of freedom for the quad models is 13-9=4). Most of the flux chi-squared contribution comes from image A being too bright. This system was also recovered by \citet{rusu2018} in a lens search using the AGN catalogue of \citet{secrest2015}.

\subsubsection{J1831+5447}
This is an NVSS radio source with a flux of 23.5mJy at 1.4GHz.

\subsubsection{J2032\MINUS 2358}
MgII absorption is seen in just one of the quasar spectra, with the emission lines appearing similar. The MgII absorption system is at z=1.642, consistent with the source's systemic redshift based on the CIII and CIV lines and the other quasar's MgII emission line. This could be a lens with magnesium absorption in the host quasar along the line of sight of just one image. If deeper imaging reveals a lens galaxy or a lensed host galaxy, then this system could be used to constrain the covering fraction of MgII.

\subsection{Lens Statistics}
To understand the limitations of our selection method, we have compiled a list of 147 previously known lensed quasars against which we can test our selection. While we report 24 new lensed quasars in this paper, our selection criteria also selected 59 previously known lensed quasars, as described in Table \ref{tab:knownlenses}.

\subsubsection{Recovering Known Lenses}
Of the 171 known lenses (including our sample), 4 lenses have 3 \Gaia detections, 43 have 2 detections, 86 have 1 detection and 38 have no detections. Of the 38 with no \Gaia detections, only 2 have bright enough images that should be detectable by \Gaia---J0941+0518 and WFI2026-4536---with the rest being optically faint mainly due to radio selection.

\begin{figure}
	\includegraphics[width=\columnwidth]{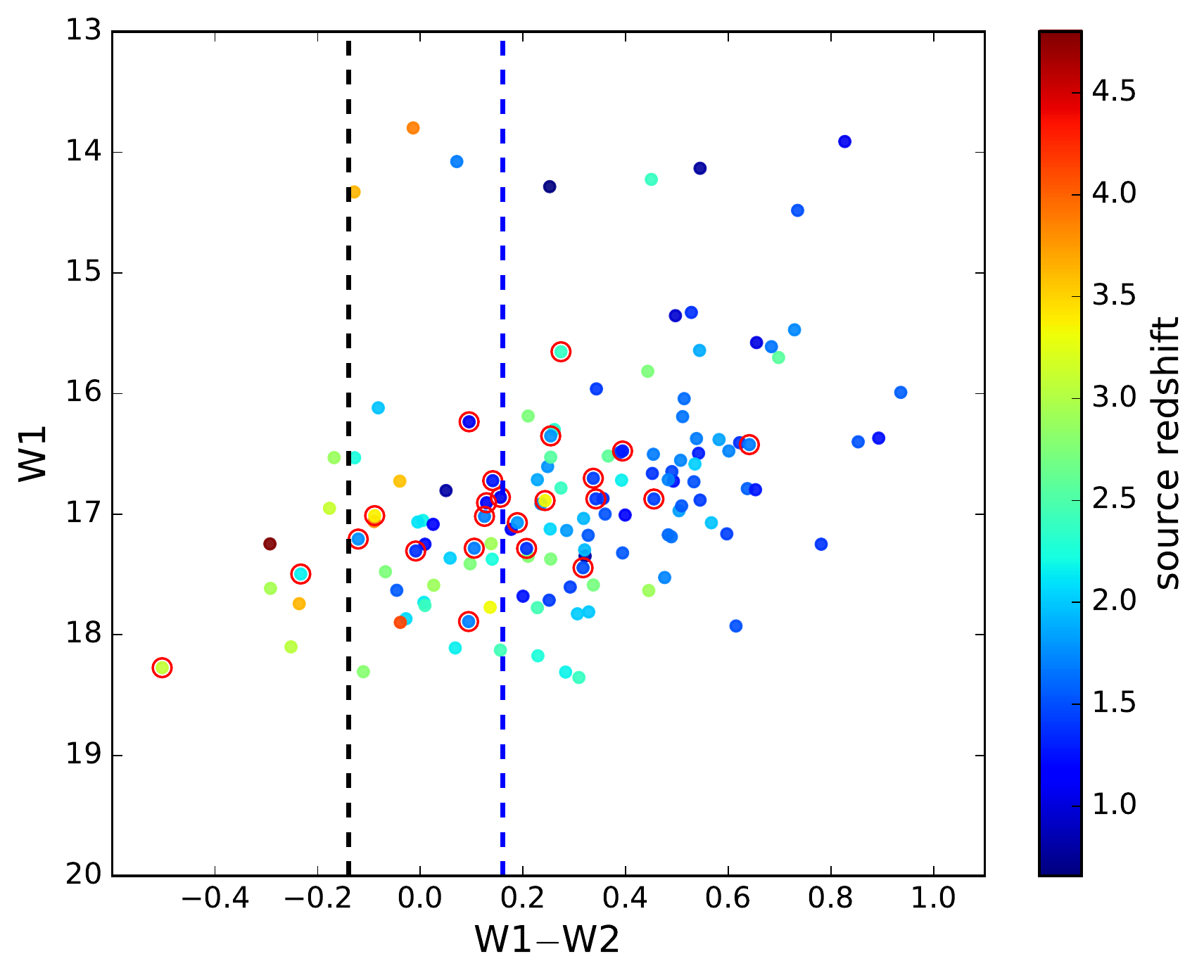}
    \caption{W1 against W1$-$W2 for all known lenses with at least one \Gaia detection and robust WISE photometry, coloured by source redshift. Our \Gaia sample are circled. Lenses circled to the left of the W1$-$W2=-0.14 black dashed line were selected with the GMM method only. The blue dashed line shows the quasar selection criterion of \citet{stern2012}.}
    \label{fig:w1w2}
\end{figure}

Applying our selection criteria (Section \ref{selection}), we recover 82 lenses (37 \Gaia multiples and 45 \Gaia singles). 23 of these lenses are quadruply imaged (quad fraction of 28$\%$), while of all known lenses the quad fraction is 43/171 (25$\%$). The lenses that we fail to select fall into three categories: separation > $4 \arcsec$, separation $\sim 1 \arcsec$, or high flux ratio doubles. We do not recover the very rare large-separation lenses because of our $4 \arcsec$ \textit{Gaia} multiple separation cut. The small-separation lenses are missed because of lack of multiple \textit{Gaia} detections \citep[see][]{lemon2017}, and in the case of just one \textit{Gaia} detection, the system is not extended enough and does not pass our morphological classifier. This latter failure also explains the high flux ratio doubles being missed, since the majority of the flux is in one PSF.

\begin{table}
	\centering
	\caption{\Gaia selection of lenses with WISE+GMM colour techniques. Numbers shown in brackets are for lenses in the Pan-STARRS footprint. 82 lenses are selected in Pan-STARRS, 23 of which are new lenses presented in this paper and the remaining 59 are known lenses.}
	\label{tab:knownlenses}
	\begin{tabular}{lc}
		\hline
         Known lensed quasars (in PS) & 171 (150) \\
         3 Gaia detections (in PS) & 4 (4)\\
         2 Gaia detections\tablefootnote{One of these 43 lenses---SBS1520+530---has one of its two detections due to a nearby star. We keep this in our sample since it would still be selected via this method.} (in PS) & 43 (34) \\
         1 Gaia detections (in PS) & 86 (77) \\
         0 Gaia detections (in PS) & 38 (35) \\
         \hline
         Multiples selected by WISE or GMM in PS & 37 \\
         Singles selected by WISE or GMM in PS & 45 \\
        \hline
	\end{tabular}
\end{table}

\subsubsection{Known versus Simulated Lenses}
Our search techniques failed to recover 33 known lenses with \Gaia detections. We understand this as the extremes of small- and high-separation images and high flux-ratio doubles. We can verify this and infer which lenses our search is missing by comparing the selected lenses to a simulated sample.

In particular we compare to the \citet{oguri2010} (OM10 hereafter) simulations which have readily listed image configurations, lensing galaxy parameters, source parameters, etc. for 15658 mock systems. We limit the entire OM10 sample to those lenses we would expect to be able to find with images brighter than the \textit{Gaia} threshold. For this criterion we use all lenses that have the faintest image (or second faintest for quads) brighter than $i$=21, leaving a catalogue of 2560 mock lenses.

\begin{figure}
	\includegraphics[width=\columnwidth]{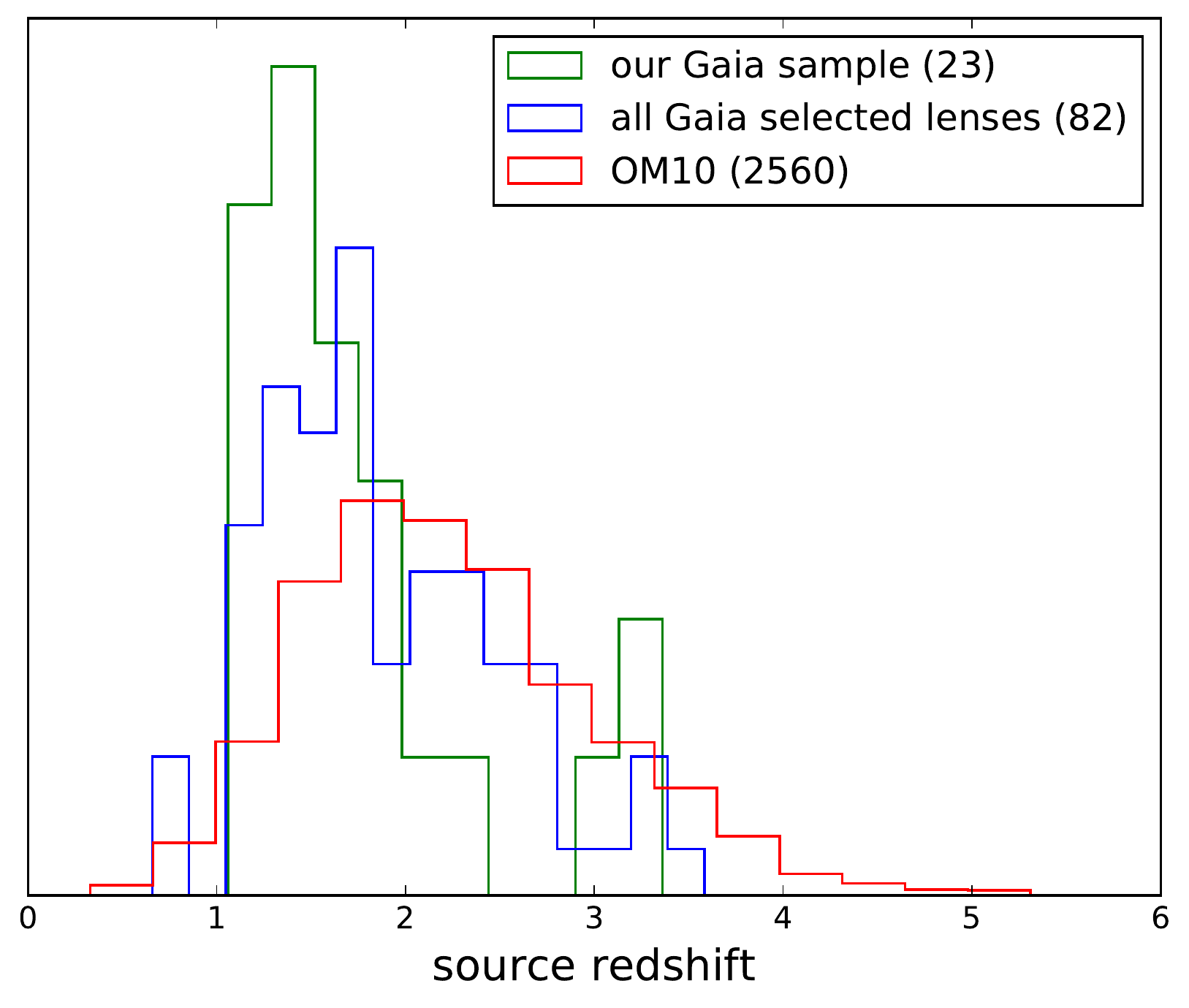}
    \caption{Source redshift distributions for our \Gaia sample, all \Gaia-selected lenses and OM10 mock lenses.}
    \label{fig:sourceredshifts}
\end{figure}

\begin{figure}
	\includegraphics[width=\columnwidth]{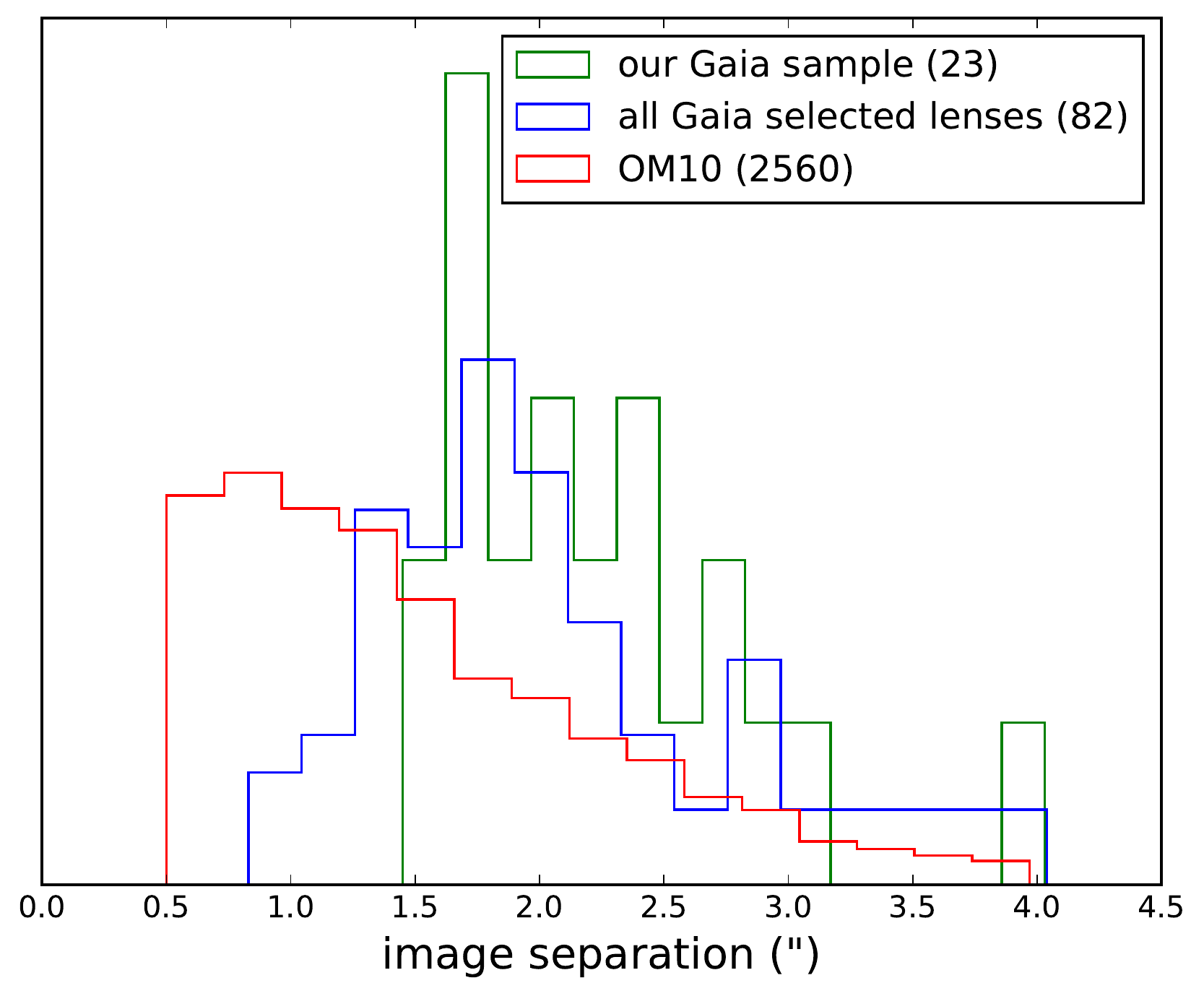}
    \caption{Image separation distributions for our \Gaia sample, all \textit{Gaia}-selected lenses and OM10 mock lenses. OM10 used a lower limit of 0\farcs5 for the image separation of their mocks and 4\arcsec \ as an upper limit.}
    \label{fig:imageseparations}
\end{figure}

Figures \ref{fig:sourceredshifts} and \ref{fig:imageseparations} show histograms of source redshifts and separations respectively for our 23 \Gaia lens sample, the 82 \textit{Gaia}-selected known lenses and the reduced OM10 sample. The first plot demonstrates the lack of known lenses at redshifts above $z=2.2$. Two reasons for this are: (i) SQLS targetted lenses below this redshift for their statistical sample and (ii) the $u$-band dropout for quasars above $z=2.7$ makes them more difficult to classify, and at higher redshifts WISE colours tend to become bluer for quasars (as in Figure \ref{fig:w1w2}). The second plot shows that small-separation lenses are being missed. This is naturally expected since these are the hardest to identify in imaging data and to target for follow-up spectroscopy.

We note that the quad fraction of the mocks is 18$\%$, compared to 25$\%$ for all known lenses, demonstrating current and previous search biases towards quads---they are more identifiable once visually inspected and more likely to be caught by search algorithms (e.g. higher likelihood of multiple \Gaia detections). 


\begin{figure}
	\includegraphics[width=\columnwidth]{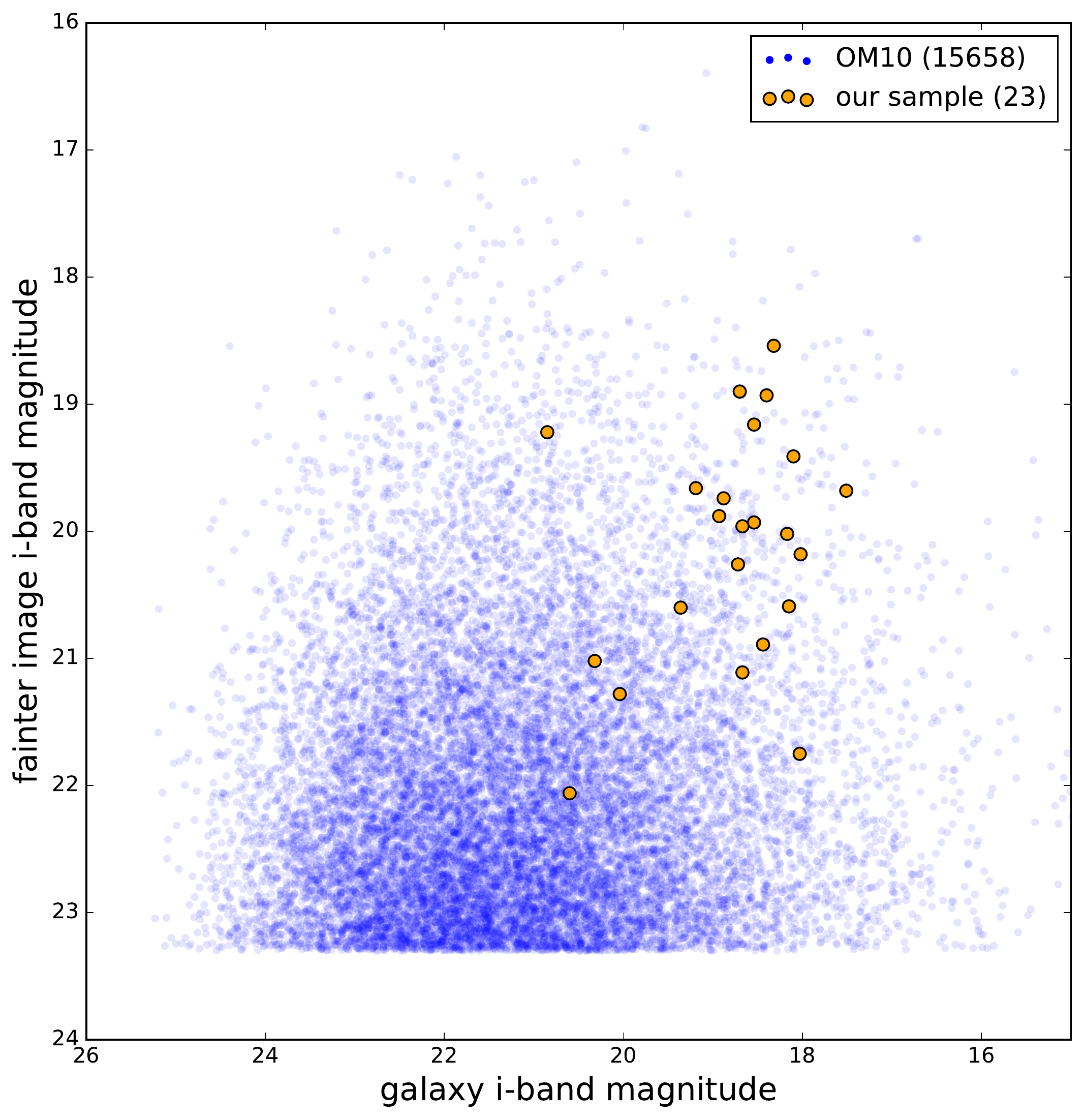}
    \caption{Fainter image (second faintest for quads) magnitudes against lensing galaxy magnitudes in the $i$-band. Yellow dots are from our modelled sample and blue is the entire OM10 simulations (no brightness limit).}
    \label{fig:imagegalmags}
\end{figure}

Figure \ref{fig:imagegalmags} shows the fainter image magnitude against galaxy magnitude in the $i$-band for our \Gaia lens sample. Overlaid are the same values for the entire OM10 mock catalogue. Naturally we are still discovering the brightest lenses on the sky, but we note that there should be many more lenses with bright images with faint galaxies, and faint images with bright galaxies (the top left and bottom right respectively). The NIQs identified in this paper can make up the former of these two classes of lens, but require deeper imaging to reveal the lens galaxy. The latter will be missed by searches requiring quasar colour selection. These lenses could be discovered by starting from photometric galaxies and requiring multiple \Gaia detections, as \mbox{\cite{lucey2017}} have successfully applied to the Pan-STARRS dataset.

\begin{table*}
	\centering
	\caption{Pan-STARRS astrometry and photometry of the lensed quasars. Magnitudes are in the AB sytem.}
	\label{tab:astrophotometry}
	\begin{tabular}{ccccccccc}
	\hline
	& component & $\alpha$ ($\arcsec$) & $\delta$ ($\arcsec$) & g & r & i & z & Y\\
	\hline
J0011\MINUS 0845 & A & 0.46 $\pm$ 0.02 & 1.24 $\pm$ 0.01 & 20.90 $\pm$ 0.01 & 20.67 $\pm$ 0.01 & 20.69 $\pm$ 0.01 & 20.64 $\pm$ 0.03 & 20.78 $\pm$ 0.05 \\
           & B & -0.48 $\pm$ 0.10 & -0.44 $\pm$ 0.10 & 21.03 $\pm$ 0.02 & 20.78 $\pm$ 0.06 & 21.02 $\pm$ 0.09 & 20.41 $\pm$ 0.23 & 20.52 $\pm$ 0.14 \\
           & G & 0.00 $\pm$ 0.05 & 0.00 $\pm$ 0.04 & 22.71 $\pm$ 0.22 & 20.85 $\pm$ 0.12 & 20.32 $\pm$ 0.09 & 20.03 $\pm$ 0.15 & 19.89 $\pm$ 0.17 \\
		\hline
J0028+0631 & A & 0.44 $\pm$ 0.01 & 2.09 $\pm$ 0.01 & 18.82 $\pm$ 0.01 & 18.75 $\pm$ 0.01 & 18.98 $\pm$ 0.01 & 19.23 $\pm$ 0.01 & 19.11 $\pm$ 0.02 \\
           & B & -0.23 $\pm$ 0.01 & -0.59 $\pm$ 0.10 & 20.65 $\pm$ 0.01 & 20.39 $\pm$ 0.02 & 20.89 $\pm$ 0.04 & 21.07 $\pm$ 0.07 & 20.73 $\pm$ 0.07 \\
           & G & 0.00 $\pm$ 0.01 & 0.00 $\pm$ 0.01 & 20.23 $\pm$ 0.02 & 19.11 $\pm$ 0.12 & 18.44 $\pm$ 0.01 & 18.19 $\pm$ 0.01 & 18.00 $\pm$ 0.02 \\
		\hline
J0030\MINUS 1525 & A & -1.09 $\pm$ 0.01 & 0.86 $\pm$ 0.01 & 19.54 $\pm$ 0.01 & 19.22 $\pm$ 0.01 & 19.07 $\pm$ 0.01 & 18.86 $\pm$ 0.01 & 18.87 $\pm$ 0.01 \\
           & B & -1.05 $\pm$ 0.19 & -0.92 $\pm$ 0.01 & 20.11 $\pm$ 0.01 & 19.78 $\pm$ 0.01 & 19.74 $\pm$ 0.01 & 19.66 $\pm$ 0.01 & 19.89 $\pm$ 0.02 \\
           & G & 0.00 $\pm$ 0.09 & 0.0 $\pm$ 0.09 & 20.58 $\pm$ 0.02 & 19.34 $\pm$ 0.012 & 18.88 $\pm$ 0.01 & 18.53 $\pm$ 0.01 & 18.37 $\pm$ 0.02 \\
		\hline
J0123\MINUS 0455 & A & 1.20 $\pm$ 0.01 & 0.93 $\pm$ 0.01 & 19.08 $\pm$ 0.01 & 19.11 $\pm$ 0.01 & 19.87 $\pm$ 0.01 & 19.67 $\pm$ 0.01 & 20.06 $\pm$ 0.03 \\
           & B & -0.43 $\pm$ 0.01 & -0.25 $\pm$ 0.01 & 19.77 $\pm$ 0.01 & 19.47 $\pm$ 0.02 & 20.59 $\pm$ 0.07 & 19.98 $\pm$ 0.07 & 20.46 $\pm$ 0.24 \\
           & G & 0.00 $\pm$ 0.01 & 0.00 $\pm$ 0.01 & 20.28 $\pm$ 0.04 & 19.12 $\pm$ 0.03 & 18.15 $\pm$ 0.02 & 18.17 $\pm$ 0.02 & 17.80 $\pm$ 0.03 \\
		\hline
J0140\MINUS 1152 & A & 0.57 $\pm$ 0.01 & 0.11 $\pm$ 0.01 & 19.54 $\pm$ 0.01 & 19.08 $\pm$ 0.01 & 18.87 $\pm$ 0.01 & 19.09 $\pm$ 0.01 & 18.92 $\pm$ 0.02 \\
           & B & -0.81 $\pm$ 0.01 & -0.30 $\pm$ 0.01 & 19.62 $\pm$ 0.01 & 19.30 $\pm$ 0.01 & 19.16 $\pm$ 0.01 & 19.23 $\pm$ 0.01 & 19.16 $\pm$ 0.02 \\
           & G & 0.00 $\pm$ 0.02 & 0.00 $\pm$ 0.01 & 20.18 $\pm$ 0.02 & 18.92 $\pm$ 0.02 & 18.53 $\pm$ 0.01 & 18.38 $\pm$ 0.03 & 17.87 $\pm$ 0.02 \\
		\hline
J0146\MINUS 1133 & A & 0.34 $\pm$ 0.01 & -1.20 $\pm$ 0.01 & 18.91 $\pm$ 0.01 & 18.74 $\pm$ 0.01 & 18.79 $\pm$ 0.01 & 18.74 $\pm$ 0.01 & 18.83 $\pm$ 0.01 \\
           & B & -0.35 $\pm$ 0.01 & 0.33 $\pm$ 0.01 & 19.17 $\pm$ 0.01 & 18.85 $\pm$ 0.01 & 18.90 $\pm$ 0.01 & 18.83 $\pm$ 0.01 & 19.12 $\pm$ 0.03 \\
           & G & 0.00 $\pm$ 0.01 & 0.00 $\pm$ 0.01 & 20.32 $\pm$ 0.03 & 19.30 $\pm$ 0.02 & 18.70 $\pm$ 0.02 & 18.52 $\pm$ 0.02 & 18.24 $\pm$ 0.02 \\
		\hline
J0235\MINUS 2433 & A & 0.48 $\pm$ 0.01 & -1.30 $\pm$ 0.01 & 19.73 $\pm$ 0.02 & 19.47 $\pm$ 0.02 & 19.58 $\pm$ 0.02 & 19.90 $\pm$ 0.03 & 19.66 $\pm$ 0.05 \\
           & B & 0.06 $\pm$ 0.01 & 0.71 $\pm$ 0.01 & 20.24 $\pm$ 0.03 & 19.93 $\pm$ 0.02 & 19.41 $\pm$ 0.02 & 20.16 $\pm$ 0.03 & 19.55 $\pm$ 0.07 \\
           & G & 0.00 $\pm$ 0.01 & 0.00 $\pm$ 0.01 & 19.43 $\pm$ 0.03 & 18.43 $\pm$ 0.02 & 18.10 $\pm$ 0.01 & 17.92 $\pm$ 0.01 & 17.59 $\pm$ 0.02 \\
		\hline
J0259\MINUS 2338 & A & -2.29 $\pm$ 0.01 & 0.19 $\pm$ 0.01 & 19.09 $\pm$ 0.02 & 19.00 $\pm$ 0.02 & 19.66 $\pm$ 0.02 & 19.20 $\pm$ 0.02 & 19.06 $\pm$ 0.09 \\
           & B & 0.59 $\pm$ 0.01 & -0.22 $\pm$ 0.01 & 19.83 $\pm$ 0.03 & 19.79 $\pm$ 0.04 & 20.26 $\pm$ 0.03 & 20.06 $\pm$ 0.05 & 19.58 $\pm$ 0.13 \\
           & G & 0.00 $\pm$ 0.01 & 0.00 $\pm$ 0.01 & 21.69 $\pm$ 0.25 & 19.44 $\pm$ 0.04 & 18.72 $\pm$ 0.02 & 18.13 $\pm$ 0.02 & 18.15 $\pm$ 0.05 \\
		\hline
J0417+3325 & A & 0.54 $\pm$ 0.01 & 1.01 $\pm$ 0.02 & 20.35 $\pm$ 0.01 & 19.64 $\pm$ 0.01 & 19.41 $\pm$ 0.01 & 19.14 $\pm$ 0.01 & 19.04 $\pm$ 0.01 \\
           & B & -0.46 $\pm$ 0.01 & -0.23 $\pm$ 0.01 & 21.10 $\pm$ 0.02 & 20.22 $\pm$ 0.02 & 19.93 $\pm$ 0.03 & 19.59 $\pm$ 0.03 & 19.39 $\pm$ 0.04 \\ 
           & G & 0.00 $\pm$ 0.01 & 0.00 $\pm$ 0.01 & 20.58 $\pm$ 0.02 & 19.33 $\pm$ 0.01 & 18.54 $\pm$ 0.01 & 18.07 $\pm$ 0.02 & 17.93 $\pm$ 0.02 \\
		\hline
J0630\MINUS 1201 & A & 0.0 $\pm$ 0.01 & 0.00 $\pm$ 0.01 & 20.50 $\pm$ 0.01 & 19.69 $\pm$ 0.01 & 19.60 $\pm$ 0.01 & 19.12 $\pm$ 0.01 & 19.14 $\pm$ 0.02 \\
           & B & 0.53 $\pm$ 0.01 & -0.58 $\pm$ 0.01 & 20.58 $\pm$ 0.02 & 19.86 $\pm$ 0.02 & 19.44 $\pm$ 0.01 & 19.18 $\pm$ 0.02 & 19.12 $\pm$ 0.02 \\
           & C & -0.31 $\pm$ 0.01 & 1.13 $\pm$ 0.01 & 20.48 $\pm$ 0.01 & 19.65 $\pm$ 0.01 & 19.44 $\pm$ 0.01 & 19.16 $\pm$ 0.01 & 19.05 $\pm$ 0.01 \\
		\hline
J0840+3550 & A & -2.04 $\pm$ 0.01 & 0.56 $\pm$ 0.01 & 20.38 $\pm$ 0.01 & 20.58 $\pm$ 0.01 & 19.92 $\pm$ 0.01 & 19.99 $\pm$ 0.01 & 20.34 $\pm$ 0.01 \\
           & B & 0.64 $\pm$ 0.04 & -0.14 $\pm$ 0.03 & 21.59 $\pm$ 0.05 & 22.06 $\pm$ 0.15 & 21.75 $\pm$ 0.20 & 21.96 $\pm$ 0.23 & 22.43 $\pm$ 0.48 \\
           & G & 0.00 $\pm$ 0.02 & 0.00 $\pm$ 0.01 & 19.48 $\pm$ 0.02 & 18.54 $\pm$ 0.01 & 18.03 $\pm$ 0.01 & 17.69 $\pm$ 0.01 & 17.65 $\pm$ 0.01 \\
		\hline
J0941+0518 & A & -3.02 $\pm$ 0.01 & -2.18 $\pm$ 0.01 & 19.39 $\pm$ 0.01 & 18.96 $\pm$ 0.01 & 18.66 $\pm$ 0.01 & 18.85 $\pm$ 0.01 & 18.76 $\pm$ 0.01 \\
           & B & 0.97 $\pm$ 0.01 & 1.46 $\pm$ 0.01 & 20.16 $\pm$ 0.01 & 19.93 $\pm$ 0.01 & 19.68 $\pm$ 0.01 & 19.86 $\pm$ 0.01 & 19.65 $\pm$ 0.02 \\
           & G & 0.0 $\pm$ 0.01 & 0.00 $\pm$ 0.01 & 19.43 $\pm$ 0.01 & 18.08 $\pm$ 0.01 & 17.51 $\pm$ 0.01 & 17.16 $\pm$ 0.01 & 16.85 $\pm$ 0.01 \\
		\hline
J0949+4208 & A & 1.04 $\pm$ 0.01 & 1.77 $\pm$ 0.09 & 19.10 $\pm$ 0.02 & 19.02 $\pm$ 0.01 & 19.61 $\pm$ 0.03 & 19.53 $\pm$ 0.02 & 19.75 $\pm$ 0.04 \\
           & B & -0.31 $\pm$ 0.01 & -0.42 $\pm$ 0.01 & 20.19 $\pm$ 0.02 & 19.99 $\pm$ 0.02 & 19.66 $\pm$ 0.02 & 19.90 $\pm$ 0.03 & 19.97 $\pm$ 0.05 \\
           & G & 0.00 $\pm$ 0.02 & 0.00 $\pm$ 0.04 & 20.87 $\pm$ 0.12 & 19.59 $\pm$ 0.04 & 19.19 $\pm$ 0.05 & 18.54 $\pm$ 0.02 & 18.45 $\pm$ 0.04 \\
		\hline
J1508+3844 & A & -0.50 $\pm$ 0.03 & 0.70 $\pm$ 0.03 & 22.18 $\pm$ 0.02 & 21.95 $\pm$ 0.04 & 21.58 $\pm$ 0.06 & 21.27 $\pm$ 0.07 & 21.09 $\pm$ 0.12 \\
           & B & 0.96 $\pm$ 0.03 & -0.19 $\pm$ 0.03 & 22.14 $\pm$ 0.02 & 21.79 $\pm$ 0.03 & 21.28 $\pm$ 0.05 & 20.93 $\pm$ 0.04 & 21.38 $\pm$ 0.11 \\
           & G & 0.00 $\pm$ 0.04 & 0.00 $\pm$ 0.05 & 22.47 $\pm$ 0.10 & 21.22 $\pm$ 0.05 & 20.04 $\pm$ 0.05 & 19.72 $\pm$ 0.04 & 19.39 $\pm$ 0.05 \\
		\hline
J1602+4526 & A & 1.99 $\pm$ 0.01 & 0.38 $\pm$ 0.01 & 19.99 $\pm$ 0.01 & 20.39 $\pm$ 0.01 & 20.20 $\pm$ 0.01 & 19.92 $\pm$ 0.01 & 20.05 $\pm$ 0.01 \\
           & B & -0.52 $\pm$ 0.02 & -0.60 $\pm$ 0.02 & 20.93 $\pm$ 0.01 & 21.19 $\pm$ 0.04 & 21.11 $\pm$ 0.03 & 20.72 $\pm$ 0.04 & 20.98 $\pm$ 0.08 \\
           & G & 0.00 $\pm$ 0.01 & 0.00 $\pm$ 0.01 & 20.67 $\pm$ 0.03 & 19.27 $\pm$ 0.02 & 18.67 $\pm$ 0.02 & 18.24 $\pm$ 0.02 & 18.04 $\pm$ 0.02 \\
		\hline
J1606\MINUS 2333 & A & -0.88 $\pm$ 0.01 & 0.30 $\pm$ 0.01 & 19.24 $\pm$ 0.01 & 19.11 $\pm$ 0.01 & 18.80 $\pm$ 0.01 & 18.88 $\pm$ 0.01 & 18.92 $\pm$ 0.05 \\
                 & B & 0.76 $\pm$ 0.01 & -0.27 $\pm$ 0.01 & 19.42 $\pm$ 0.01 & 19.33 $\pm$ 0.01 & 18.99 $\pm$ 0.01 & 19.08 $\pm$ 0.02 & 19.12 $\pm$ 0.06 \\
                 & C & -0.07 $\pm$ 0.01 & -0.60 $\pm$ 0.01 & 19.79 $\pm$ 0.01 & 19.51 $\pm$ 0.02 & 19.22 $\pm$ 0.01 & 19.23 $\pm$ 0.02 & 19.27 $\pm$ 0.06 \\
                 & D & 0.24 $\pm$ 0.01 & 0.48 $\pm$ 0.01 & 20.23 $\pm$ 0.03 & 19.95 $\pm$ 0.03 & 19.46 $\pm$ 0.02 & 19.59 $\pm$ 0.02 & 19.66 $\pm$ 0.07 \\
                 & G & 0.00 $\pm$ 0.05 & 0.00 $\pm$ 0.05 & 21.67 $\pm$ 0.29 & 21.40 $\pm$ 0.25 & 20.85 $\pm$ 0.17 & 21.08 $\pm$ 0.29 & 19.75 $\pm$ 0.19 \\
		\hline
J1640+1045 & A & -0.37 $\pm$ 0.01 & -0.36 $\pm$ 0.01 & 18.69 $\pm$ 0.01 & 18.73 $\pm$ 0.01 & 18.49 $\pm$ 0.01 & 18.28 $\pm$ 0.01 & 18.34 $\pm$ 0.01 \\
           & B & 1.44 $\pm$ 0.01 & 0.92 $\pm$ 0.01 & 20.31 $\pm$ 0.01 & 20.34 $\pm$ 0.01 & 19.96 $\pm$ 0.01 & 19.89 $\pm$ 0.01 & 19.87 $\pm$ 0.01 \\
           & G & 0.00 $\pm$ 0.01 & 0.00 $\pm$ 0.01 & 20.66 $\pm$ 0.04 & 19.25 $\pm$ 0.02 & 18.67 $\pm$ 0.01 & 18.66 $\pm$ 0.02 & 18.29 $\pm$ 0.02 \\
		\hline
	\end{tabular}
\end{table*}

\begin{table*}
	\centering
	\begin{tabular}{ccccccccc}
	\hline
J1709+3828 & A & 0.75 $\pm$ 0.01 & -0.44 $\pm$ 0.02 & 21.49 $\pm$ 0.02 & 20.68 $\pm$ 0.01 & 20.70 $\pm$ 0.03 & 20.89 $\pm$ 0.05 & 21.29 $\pm$ 0.23 \\
           & B & -0.05 $\pm$ 0.01 & 1.06 $\pm$ 0.01 & 21.21 $\pm$ 0.01 & 20.65 $\pm$ 0.01 & 20.60 $\pm$ 0.02 & 20.91 $\pm$ 0.03 & 20.72 $\pm$ 0.07 \\
           & G & 0.00 $\pm$ 0.02 & 0.00 $\pm$ 0.02 & 21.57 $\pm$ 0.04 & 20.06 $\pm$ 0.02 & 19.36 $\pm$ 0.02 & 18.98 $\pm$ 0.02 & 18.79 $\pm$ 0.03 \\
		\hline
J1710+4332 & A & 0.29 $\pm$ 0.01 & 1.67 $\pm$ 0.01 & 20.89 $\pm$ 0.01 & 21.09 $\pm$ 0.01 & 21.43 $\pm$ 0.01 & 21.20 $\pm$ 0.02 & 20.20 $\pm$ 0.02 \\
           & B & -0.17 $\pm$ 0.02 & -0.72 $\pm$ 0.02 & 21.89 $\pm$ 0.01 & 21.99 $\pm$ 0.04 & 22.06 $\pm$ 0.07 & 21.34 $\pm$ 0.12 & 20.18 $\pm$ 0.15 \\
           & G & 0.00 $\pm$ 0.05 & 0.00 $\pm$ 0.06 & --- & 21.60 $\pm$ 0.08 & 20.60 $\pm$ 0.09 & 20.22 $\pm$ 0.09 & 20.38 $\pm$ 0.18 \\
		\hline
J1721+8842 & A & -0.54 $\pm$ 0.01 & -1.80 $\pm$ 0.01 & 19.45 $\pm$ 0.01 & 19.50 $\pm$ 0.02 & 19.14 $\pm$ 0.03 & 19.39 $\pm$ 0.02 & 19.14 $\pm$ 0.05 \\
           & B & -2.13 $\pm$ 0.01 & -0.71 $\pm$ 0.01 & 20.57 $\pm$ 0.02 & 20.45 $\pm$ 0.02 & 20.18 $\pm$ 0.03 & 20.15 $\pm$ 0.03 & 19.88 $\pm$ 0.05 \\
           & C & 1.88 $\pm$ 0.01 & -1.10 $\pm$ 0.01 & 20.13 $\pm$ 0.01 & 20.04 $\pm$ 0.02 & 19.90 $\pm$ 0.03 & 19.91 $\pm$ 0.02 & 19.71 $\pm$ 0.05 \\
           & D & 0.06 $\pm$ 0.02 & 1.69 $\pm$ 0.02 & 21.69 $\pm$ 0.04 & 21.43 $\pm$ 0.03 & 21.02 $\pm$ 0.05 & 21.18 $\pm$ 0.04 & 21.11 $\pm$ 0.08 \\
           & G & 0.00 $\pm$ 0.01 & 0.00 $\pm$ 0.02 & 18.89 $\pm$ 0.02 & 18.47 $\pm$ 0.02 & 18.02 $\pm$ 0.03 & 17.97 $\pm$ 0.02 & 18.01 $\pm$ 0.04 \\
		\hline
J1831+5447 & A & 1.47 $\pm$ 0.01 & 1.10 $\pm$ 0.01 & 19.32 $\pm$ 0.01 & 19.14 $\pm$ 0.03 & 19.22 $\pm$ 0.02 & 19.08 $\pm$ 0.04 & 19.21 $\pm$ 0.06 \\
           & B & -0.37 $\pm$ 0.01 & -0.27 $\pm$ 0.01 & 20.55 $\pm$ 0.03 & 20.07 $\pm$ 0.04 & 20.02 $\pm$ 0.03 & 19.58 $\pm$ 0.05 & 19.76 $\pm$ 0.08 \\
           & G & 0.00 $\pm$ 0.02 & 0.00 $\pm$ 0.01 & 20.21 $\pm$ 0.07 & 18.95 $\pm$ 0.04 & 18.17 $\pm$ 0.01 & 17.76 $\pm$ 0.02 & 17.45 $\pm$ 0.03 \\
		\hline
J2124+1632 & A & -0.04 $\pm$ 0.01 & 2.55 $\pm$ 0.01 & 18.99 $\pm$ 0.01 & 18.66 $\pm$ 0.01 & 18.80 $\pm$ 0.02 & 18.93 $\pm$ 0.03 & 19.08 $\pm$ 0.06 \\
           & B & 0.13 $\pm$ 0.01 & -0.46 $\pm$ 0.01 & 19.24 $\pm$ 0.02 & 19.16 $\pm$ 0.02 & 18.93 $\pm$ 0.03 & 19.34 $\pm$ 0.03 & 19.34 $\pm$ 0.08 \\
           & G & 0.00 $\pm$ 0.01 & 0.00 $\pm$ 0.01 & 20.44 $\pm$ 0.10 & 18.88 $\pm$ 0.04 & 18.40 $\pm$ 0.04 & 18.03 $\pm$ 0.04 & 17.89 $\pm$ 0.05 \\
		\hline
J2305+3714 & A & 1.18 $\pm$ 0.01 & -0.83 $\pm$ 0.01 & 17.96 $\pm$ 0.01 & 17.68 $\pm$ 0.01 & 17.34 $\pm$ 0.01 & 17.30 $\pm$ 0.01 & 17.28 $\pm$ 0.02 \\
           & B & -0.25 $\pm$ 0.01 & 0.84 $\pm$ 0.01 & 18.94 $\pm$ 0.01 & 18.99 $\pm$ 0.01 & 18.54 $\pm$ 0.01 & 18.51 $\pm$ 0.01 & 18.46 $\pm$ 0.02 \\
           & G & 0.00 $\pm$ 0.02 & 0.00 $\pm$ 0.02 & 22.14 $\pm$ 0.43 & 18.59 $\pm$ 0.04 & 18.32 $\pm$ 0.03 & 18.05 $\pm$ 0.03 & 17.81 $\pm$ 0.05 \\
		\hline        
J2332\MINUS 1852 & A & 1.35 $\pm$ 0.01 & 0.43 $\pm$ 0.01 & 20.27 $\pm$ 0.01 & 20.24 $\pm$ 0.01 & 19.24 $\pm$ 0.01 & 19.68 $\pm$ 0.01 & 19.15 $\pm$ 0.01 \\
           & B & -0.43 $\pm$ 0.01 & -0.41 $\pm$ 0.01 & 20.59 $\pm$ 0.01 & 20.59 $\pm$ 0.01 & 19.88 $\pm$ 0.03 & 20.21 $\pm$ 0.03 & 19.44 $\pm$ 0.03 \\
           & G & 0.00 $\pm$ 0.02 & 0.00 $\pm$ 0.02 & --- & 21.01 $\pm$ 0.05 & 18.93 $\pm$ 0.04 & 19.40 $\pm$ 0.04 & 18.71 $\pm$ 0.04 \\
		\hline
	\end{tabular}
\end{table*}

\section{Conclusions} \label{conclusions}
We have presented the discovery of 24 gravitationally lensed quasars; 13 are discovered by having multiple \Gaia components, 10 as \Gaia detections near morphological galaxies and 1 from an emission line identification in an SDSS galaxy spectrum. All of these systems have well-resolved images in ground-based imaging and are bright, lending themselves to monitoring and subsequent time delay cosmography studies \mbox{\citep{treu2016}}. Our sample includes 4 quadruply-imaged lenses and one double lens which shows a blue Einstein ring in deep CFHT data. One quad, J0030-1525, requires high resolution imaging to verify our understanding that one image is highly demagnified, by a factor of $\sim$ 14. Four of our lenses---J0140-1152, J0146-1133, J0235-2433, J0259-2338---have been independently identified using a complementary \textit{Gaia} and WISE selection by \citet{agnellovst2017} and \citet{agnello2017des}.

Comparing our sample of lensed quasars to simulations, we show that we are not sensitive to arcsecond-separation lenses and those at redshifts beyond the $u$-band dropout. This is due to colour selection and the bias from visually inspecting a large number of candidates. This biases our sample towards systems with bright lensing galaxies and quads, and away from small-separation lenses and bright doubles that are often confused for stars. The second \Gaia data release on 25 April 2018 will not only help increase the completeness of lens searches through detection of more lensed quasar images and colour information from the blue and red photometers, but also drastically reduce the number of candidates that must be inspected. Further data releases including proper motion information will further remove contaminants.

\mbox{\citet{oguri2010}} predict several thousand lensed quasars to be detectable using LSST (Large Synoptic Survey Telescope), however to confirm such large numbers of lenses to become useful astrophysical and cosmological tools, we must ensure we select lenses efficiently and in a complete manner. We have demonstrated that combining \textit{Gaia}, WISE, SDSS and Pan-STARRS data with pixel-based modelling can efficiently select new lenses---a 57$\%$ success rate (including binaries).

\section*{Acknowledgements}
CAL and RGM would like to thank the STFC, and MWA acknowledges support from the STFC in the form of an Ernest Rutherford Fellowship. This work has made use of data from the European Space Agency (ESA) mission {\it Gaia} (\mbox{\url{https://www.cosmos.esa.int/gaia}}), processed by the {\it Gaia} Data Processing and Analysis Consortium (DPAC, \mbox{\url{https://www.cosmos.esa.int/web/gaia/dpac/consortium}}).
Funding for the DPAC has been provided by national institutions, in particular the institutions participating in the {\it Gaia} Multilateral Agreement. The Pan-STARRS1 Surveys (PS1) and the PS1 public science archive have been made possible through contributions by the Institute for Astronomy, the University of Hawaii, the Pan-STARRS Project Office, the Max-Planck Society and its participating institutes, the Max Planck Institute for Astronomy, Heidelberg and the Max Planck Institute for Extraterrestrial Physics, Garching, The Johns Hopkins University, Durham University, the University of Edinburgh, the Queen's University Belfast, the Harvard-Smithsonian Center for Astrophysics, the Las Cumbres Observatory Global Telescope Network Incorporated, the National Central University of Taiwan, the Space Telescope Science Institute, the National Aeronautics and Space Administration under Grant No. NNX08AR22G issued through the Planetary Science Division of the NASA Science Mission Directorate, the National Science Foundation Grant No. AST-1238877, the University of Maryland, Eotvos Lorand University (ELTE), the Los Alamos National Laboratory, and the Gordon and Betty Moore Foundation. This publication makes use of data products from the Wide-field Infrared Survey Explorer, which is a joint project of the University of California, Los Angeles, and the Jet Propulsion Laboratory/California Institute of Technology, funded by the National Aeronautics and Space Administration.


\bibliographystyle{mnras}
\bibliography{papers} 

\begin{thebibliography}{}
\makeatletter
\relax
\def\mn@urlcharsother{\let\do\@makeother \do\$\do\&\do\#\do\^\do\_\do\%\do\~}
\def\mn@doi{\begingroup\mn@urlcharsother \@ifnextchar [ {\mn@doi@}
  {\mn@doi@[]}}
\def\mn@doi@[#1]#2{\def\@tempa{#1}\ifx\@tempa\@empty \href
  {http://dx.doi.org/#2} {doi:#2}\else \href {http://dx.doi.org/#2} {#1}\fi
  \endgroup}
\def\mn@eprint#1#2{\mn@eprint@#1:#2::\@nil}
\def\mn@eprint@arXiv#1{\href {http://arxiv.org/abs/#1} {{\tt arXiv:#1}}}
\def\mn@eprint@dblp#1{\href {http://dblp.uni-trier.de/rec/bibtex/#1.xml}
  {dblp:#1}}
\def\mn@eprint@#1:#2:#3:#4\@nil{\def\@tempa {#1}\def\@tempb {#2}\def\@tempc
  {#3}\ifx \@tempc \@empty \let \@tempc \@tempb \let \@tempb \@tempa \fi \ifx
  \@tempb \@empty \def\@tempb {arXiv}\fi \@ifundefined
  {mn@eprint@\@tempb}{\@tempb:\@tempc}{\expandafter \expandafter \csname
  mn@eprint@\@tempb\endcsname \expandafter{\@tempc}}}

\bibitem[\protect\citeauthoryear{{Abbott} et~al.,}{{Abbott}
  et~al.}{2018}]{desdatarelease}
{Abbott} T.~M.~C.,  et~al., 2018, preprint, \href
  {http://adsabs.harvard.edu/abs/2018arXiv180103181A} {} (\mn@eprint {arXiv}
  {1801.03181})

\bibitem[\protect\citeauthoryear{{Agnello}}{{Agnello}}{2017}]{agnellogaia}
{Agnello} A.,  2017, \mn@doi [\mnras] {10.1093/mnras/stx1650}, \href
  {http://adsabs.harvard.edu/abs/2017MNRAS.471.2013A} {471, 2013}

\bibitem[\protect\citeauthoryear{{Agnello} et~al.,}{{Agnello}
  et~al.}{2017}]{agnello2017des}
{Agnello} A.,  et~al., 2017, preprint, \href
  {http://adsabs.harvard.edu/abs/2017arXiv171103971A} {} (\mn@eprint {arXiv}
  {1711.03971})

\bibitem[\protect\citeauthoryear{{Agnello}, {Grillo}, {Jones}, {Treu},
  {Bonamigo}  \& {Suyu}}{{Agnello} et~al.}{2018a}]{agnello2018}
{Agnello} A.,  {Grillo} C.,  {Jones} T.,  {Treu} T.,  {Bonamigo} M.,   {Suyu}
  S.~H.,  2018a, \mn@doi [\mnras] {10.1093/mnras/stx2950}, \href
  {http://adsabs.harvard.edu/abs/2018MNRAS.474.3391A} {474, 3391}

\bibitem[\protect\citeauthoryear{{Agnello} et~al.,}{{Agnello}
  et~al.}{2018b}]{agnellovst2017}
{Agnello} A.,  et~al., 2018b, \mn@doi [\mnras] {10.1093/mnras/stx3226}, \href
  {http://adsabs.harvard.edu/abs/2018MNRAS.475.2086A} {475, 2086}

\bibitem[\protect\citeauthoryear{{Altschuler}}{{Altschuler}}{1986}]{altschuler1986}
{Altschuler} D.~R.,  1986, \aaps, \href
  {http://adsabs.harvard.edu/abs/1986A%26AS...65..267A} {65, 267}

\bibitem[\protect\citeauthoryear{{Arenou} et~al.,}{{Arenou}
  et~al.}{2017}]{arenou2017}
{Arenou} F.,  et~al., 2017, preprint (\mn@eprint {arXiv} {1701.00292})

\bibitem[\protect\citeauthoryear{{Auger}}{{Auger}}{2008}]{auger2008}
{Auger} M.~W.,  2008, \mn@doi [\mnras] {10.1111/j.1745-3933.2007.00408.x},
  \href {http://adsabs.harvard.edu/abs/2008MNRAS.383L..40A} {383, L40}

\bibitem[\protect\citeauthoryear{{Auger}, {Treu}, {Bolton}, {Gavazzi},
  {Koopmans}, {Marshall}, {Moustakas}  \& {Burles}}{{Auger}
  et~al.}{2010}]{auger2010}
{Auger} M.~W.,  {Treu} T.,  {Bolton} A.~S.,  {Gavazzi} R.,  {Koopmans}
  L.~V.~E.,  {Marshall} P.~J.,  {Moustakas} L.~A.,   {Burles} S.,  2010,
  \mn@doi [\apj] {10.1088/0004-637X/724/1/511}, \href
  {http://adsabs.harvard.edu/abs/2010ApJ...724..511A} {724, 511}

\bibitem[\protect\citeauthoryear{{Bate}, {Floyd}, {Webster}  \&
  {Wyithe}}{{Bate} et~al.}{2008}]{bate2008}
{Bate} N.~F.,  {Floyd} D.~J.~E.,  {Webster} R.~L.,   {Wyithe} J.~S.~B.,  2008,
  \mn@doi [\mnras] {10.1111/j.1365-2966.2008.14020.x}, 391, 1955

\bibitem[\protect\citeauthoryear{{Bate}, {Floyd}, {Webster}  \&
  {Wyithe}}{{Bate} et~al.}{2011}]{bate2011}
{Bate} N.~F.,  {Floyd} D.~J.~E.,  {Webster} R.~L.,   {Wyithe} J.~S.~B.,  2011,
  \mn@doi [\apj] {10.1088/0004-637X/731/1/71}, 731, 71

\bibitem[\protect\citeauthoryear{{Bayliss} et~al.,}{{Bayliss}
  et~al.}{2017}]{bayliss2017}
{Bayliss} M.~B.,  et~al., 2017, preprint, \href
  {http://adsabs.harvard.edu/abs/2017arXiv170800453B} {} (\mn@eprint {arXiv}
  {1708.00453})

\bibitem[\protect\citeauthoryear{{Belokurov}, {Erkal}, {Deason}, {Koposov}, {De
  Angeli}, {Wyn Evans}, {Fraternali}  \& {Mackey}}{{Belokurov}
  et~al.}{2016}]{belokurov2016}
{Belokurov} V.,  {Erkal} D.,  {Deason} A.~J.,  {Koposov} S.~E.,  {De Angeli}
  F.,  {Wyn Evans} D.,  {Fraternali} F.,   {Mackey} D.,  2016, \mn@doi [\mnras]
  {10.1093/mnras/stw3357}

\bibitem[\protect\citeauthoryear{{Bonvin} et~al.,}{{Bonvin}
  et~al.}{2017}]{bonvin2016}
{Bonvin} V.,  et~al., 2017, \mn@doi [\mnras] {10.1093/mnras/stw3006}, 465, 4914

\bibitem[\protect\citeauthoryear{{Chambers} et~al.,}{{Chambers}
  et~al.}{2016}]{chambers2016}
{Chambers} K.~C.,  et~al., 2016, preprint (\mn@eprint {arXiv} {1612.05560})

\bibitem[\protect\citeauthoryear{{Colla} et~al.,}{{Colla}
  et~al.}{1970}]{colla1970}
{Colla} G.,  et~al., 1970, \aaps, \href
  {http://adsabs.harvard.edu/abs/1970A%26AS....1..281C} {1, 281}

\bibitem[\protect\citeauthoryear{{Condon}, {Cotton}, {Greisen}, {Yin},
  {Perley}, {Taylor}  \& {Broderick}}{{Condon} et~al.}{1998}]{condon1998}
{Condon} J.~J.,  {Cotton} W.~D.,  {Greisen} E.~W.,  {Yin} Q.~F.,  {Perley}
  R.~A.,  {Taylor} G.~B.,   {Broderick} J.~J.,  1998, \mn@doi [\aj]
  {10.1086/300337}, \href {http://adsabs.harvard.edu/abs/1998AJ....115.1693C}
  {115, 1693}

\bibitem[\protect\citeauthoryear{{Davis}}{{Davis}}{1971}]{davis1971}
{Davis} M.~M.,  1971, \mn@doi [\aj] {10.1086/111211}, \href
  {http://adsabs.harvard.edu/abs/1971AJ.....76..980D} {76, 980}

\bibitem[\protect\citeauthoryear{{Ding} et~al.,}{{Ding}
  et~al.}{2017}]{ding2017}
{Ding} X.,  et~al., 2017, preprint (\mn@eprint {arXiv} {1703.02041})

\bibitem[\protect\citeauthoryear{{Dobke}, {King}  \& {Fellhauer}}{{Dobke}
  et~al.}{2007}]{dobke2007}
{Dobke} B.~M.,  {King} L.~J.,   {Fellhauer} M.,  2007, \mn@doi [\mnras]
  {10.1111/j.1365-2966.2007.11683.x}, \href
  {http://adsabs.harvard.edu/abs/2007MNRAS.377.1503D} {377, 1503}

\bibitem[\protect\citeauthoryear{{Fabricius} et~al.,}{{Fabricius}
  et~al.}{2016}]{fabricius2016}
{Fabricius} C.,  et~al., 2016, \mn@doi [\aap] {10.1051/0004-6361/201628643},
  595, A3

\bibitem[\protect\citeauthoryear{{Foreman-Mackey}, {Hogg}, {Lang}  \&
  {Goodman}}{{Foreman-Mackey} et~al.}{2013}]{foreman-mackey2013}
{Foreman-Mackey} D.,  {Hogg} D.~W.,  {Lang} D.,   {Goodman} J.,  2013, \mn@doi
  [\pasp] {10.1086/670067}, \href
  {http://adsabs.harvard.edu/abs/2013PASP..125..306F} {125, 306}

\bibitem[\protect\citeauthoryear{{Gaia Collaboration} et~al.,}{{Gaia
  Collaboration} et~al.}{2016a}]{prusti2016}
{Gaia Collaboration} et~al., 2016a, \mn@doi [\aap]
  {10.1051/0004-6361/201629272}, 595, A1

\bibitem[\protect\citeauthoryear{{Gaia Collaboration} et~al.,}{{Gaia
  Collaboration} et~al.}{2016b}]{brown2016}
{Gaia Collaboration} et~al., 2016b, \mn@doi [\aap]
  {10.1051/0004-6361/201629512}, \href
  {http://adsabs.harvard.edu/abs/2016A%26A...595A...2G} {595, A2}

\bibitem[\protect\citeauthoryear{{Inada} et~al.,}{{Inada}
  et~al.}{2010}]{inada2010}
{Inada} N.,  et~al., 2010, \mn@doi [\aj] {10.1088/0004-6256/140/2/403}, 140,
  403

\bibitem[\protect\citeauthoryear{{Jarrett} et~al.,}{{Jarrett}
  et~al.}{2011}]{jarrett2011}
{Jarrett} T.~H.,  et~al., 2011, \mn@doi [\apj] {10.1088/0004-637X/735/2/112},
  \href {http://adsabs.harvard.edu/abs/2011ApJ...735..112J} {735, 112}

\bibitem[\protect\citeauthoryear{{Jim{\'e}nez-Vicente}, {Mediavilla},
  {Kochanek}  \& {Mu{\~n}oz}}{{Jim{\'e}nez-Vicente} et~al.}{2015}]{jimenez2015}
{Jim{\'e}nez-Vicente} J.,  {Mediavilla} E.,  {Kochanek} C.~S.,   {Mu{\~n}oz}
  J.~A.,  2015, \mn@doi [\apj] {10.1088/0004-637X/799/2/149}, 799, 149

\bibitem[\protect\citeauthoryear{{Keeton}}{{Keeton}}{2001}]{keeton2001}
{Keeton} C.~R.,  2001, ArXiv Astrophysics e-prints, \href
  {http://adsabs.harvard.edu/abs/2001astro.ph..2340K} {}

\bibitem[\protect\citeauthoryear{{Koposov}, {Belokurov}  \&
  {Torrealba}}{{Koposov} et~al.}{2017}]{koposov2017}
{Koposov} S.~E.,  {Belokurov} V.,   {Torrealba} G.,  2017, preprint (\mn@eprint
  {arXiv} {1702.01122})

\bibitem[\protect\citeauthoryear{{Lemon}, {Auger}, {McMahon}  \&
  {Koposov}}{{Lemon} et~al.}{2017}]{lemon2017}
{Lemon} C.~A.,  {Auger} M.~W.,  {McMahon} R.~G.,   {Koposov} S.~E.,  2017,
  \mn@doi [\mnras] {10.1093/mnras/stx2094}, \href
  {http://adsabs.harvard.edu/abs/2017MNRAS.472.5023L} {472, 5023}

\bibitem[\protect\citeauthoryear{{Lindegren}, {Lammers}, {Hobbs}, {O'Mullane},
  {Bastian}  \& {Hern{\'a}ndez}}{{Lindegren} et~al.}{2012}]{lindegren2012}
{Lindegren} L.,  {Lammers} U.,  {Hobbs} D.,  {O'Mullane} W.,  {Bastian} U.,
  {Hern{\'a}ndez} J.,  2012, \mn@doi [\aap] {10.1051/0004-6361/201117905}, 538,
  A78

\bibitem[\protect\citeauthoryear{{Lindegren} et~al.,}{{Lindegren}
  et~al.}{2016}]{lindegren2016}
{Lindegren} L.,  et~al., 2016, \mn@doi [\aap] {10.1051/0004-6361/201628714},
  \href {http://adsabs.harvard.edu/abs/2016A%26A...595A...4L} {595, A4}

\bibitem[\protect\citeauthoryear{{Lucey}, {Schechter}, {Smith}  \&
  {Anguita}}{{Lucey} et~al.}{2017}]{lucey2017}
{Lucey} J.~R.,  {Schechter} P.~L.,  {Smith} R.~J.,   {Anguita} T.,  2017,
  preprint, \href {http://adsabs.harvard.edu/abs/2017arXiv171102674L} {}
  (\mn@eprint {arXiv} {1711.02674})

\bibitem[\protect\citeauthoryear{{Mediavilla} et~al.,}{{Mediavilla}
  et~al.}{2009}]{mediavilla2009}
{Mediavilla} E.,  et~al., 2009, \mn@doi [\apj] {10.1088/0004-637X/706/2/1451},
  \href {http://adsabs.harvard.edu/abs/2009ApJ...706.1451M} {706, 1451}

\bibitem[\protect\citeauthoryear{{Moffat}}{{Moffat}}{1969}]{moffat1969}
{Moffat} A.~F.~J.,  1969, \aap, \href
  {http://adsabs.harvard.edu/abs/1969A%26A.....3..455M} {3, 455}

\bibitem[\protect\citeauthoryear{{Oguri} \& {Marshall}}{{Oguri} \&
  {Marshall}}{2010}]{oguri2010}
{Oguri} M.,  {Marshall} P.~J.,  2010, \mn@doi [\mnras]
  {10.1111/j.1365-2966.2010.16639.x}, 405, 2579

\bibitem[\protect\citeauthoryear{{Oguri} et~al.,}{{Oguri}
  et~al.}{2006}]{oguri2006}
{Oguri} M.,  et~al., 2006, \mn@doi [\aj] {10.1086/506019}, \href
  {http://adsabs.harvard.edu/abs/2006AJ....132..999O} {132, 999}

\bibitem[\protect\citeauthoryear{{Ostrovski} et~al.,}{{Ostrovski}
  et~al.}{2017}]{ostrovski2016}
{Ostrovski} F.,  et~al., 2017, \mn@doi [\mnras] {10.1093/mnras/stw2958}, 465,
  4325

\bibitem[\protect\citeauthoryear{{Ostrovski} et~al.,}{{Ostrovski}
  et~al.}{2018}]{ostrovski2017}
{Ostrovski} F.,  et~al., 2018, \mn@doi [\mnras] {10.1093/mnrasl/slx173}, \href
  {http://adsabs.harvard.edu/abs/2018MNRAS.473L.116O} {473, L116}

\bibitem[\protect\citeauthoryear{{Peng}, {Impey}, {Rix}, {Kochanek}, {Keeton},
  {Falco}, {Leh{\'a}r}  \& {McLeod}}{{Peng} et~al.}{2006}]{peng2006}
{Peng} C.~Y.,  {Impey} C.~D.,  {Rix} H.-W.,  {Kochanek} C.~S.,  {Keeton} C.~R.,
   {Falco} E.~E.,  {Leh{\'a}r} J.,   {McLeod} B.~A.,  2006, \mn@doi [\apj]
  {10.1086/506266}, \href {http://adsabs.harvard.edu/abs/2006ApJ...649..616P}
  {649, 616}

\bibitem[\protect\citeauthoryear{{Planck Collaboration} et~al.,}{{Planck
  Collaboration} et~al.}{2016}]{planck2016}
{Planck Collaboration} et~al., 2016, \mn@doi [\aap]
  {10.1051/0004-6361/201525830}, \href
  {http://adsabs.harvard.edu/abs/2016A%26A...594A..13P} {594, A13}

\bibitem[\protect\citeauthoryear{{Pooley}, {Blackburne}, {Rappaport}  \&
  {Schechter}}{{Pooley} et~al.}{2007}]{pooley2007}
{Pooley} D.,  {Blackburne} J.~A.,  {Rappaport} S.,   {Schechter} P.~L.,  2007,
  \mn@doi [\apj] {10.1086/512115}, \href
  {http://adsabs.harvard.edu/abs/2007ApJ...661...19P} {661, 19}

\bibitem[\protect\citeauthoryear{{Rauch} \& {Blandford}}{{Rauch} \&
  {Blandford}}{1991}]{rauch1991}
{Rauch} K.~P.,  {Blandford} R.~D.,  1991, \mn@doi [\apjl] {10.1086/186191},
  \href {http://adsabs.harvard.edu/abs/1991ApJ...381L..39R} {381, L39}

\bibitem[\protect\citeauthoryear{{Refsdal}}{{Refsdal}}{1966}]{refsdal1966}
{Refsdal} S.,  1966, \mn@doi [\mnras] {10.1093/mnras/132.1.101}, 132, 101

\bibitem[\protect\citeauthoryear{{Riess} et~al.,}{{Riess}
  et~al.}{2016}]{riess2016}
{Riess} A.~G.,  et~al., 2016, \mn@doi [\apj] {10.3847/0004-637X/826/1/56},
  \href {http://adsabs.harvard.edu/abs/2016ApJ...826...56R} {826, 56}

\bibitem[\protect\citeauthoryear{{Rusu}, {Berghea}, {Fassnacht}, {More},
  {Seman}  \& {Nelson}}{{Rusu} et~al.}{2018}]{rusu2018}
{Rusu} C.~E.,  {Berghea} C.~T.,  {Fassnacht} C.~D.,  {More} A.,  {Seman} E.,
  {Nelson} G.~J.,  2018, preprint, \href
  {http://adsabs.harvard.edu/abs/2018arXiv180307175R} {} (\mn@eprint {arXiv}
  {1803.07175})

\bibitem[\protect\citeauthoryear{{Schechter} \& {Wambsganss}}{{Schechter} \&
  {Wambsganss}}{2002}]{schechter2002}
{Schechter} P.~L.,  {Wambsganss} J.,  2002, \mn@doi [\apj] {10.1086/343856},
  \href {http://adsabs.harvard.edu/abs/2002ApJ...580..685S} {580, 685}

\bibitem[\protect\citeauthoryear{{Schechter} \& {Wambsganss}}{{Schechter} \&
  {Wambsganss}}{2004}]{schechter2004}
{Schechter} P.~L.,  {Wambsganss} J.,  2004, in {Ryder} S.,  {Pisano} D.,
  {Walker} M.,   {Freeman} K.,  eds,  IAU Symposium Vol. 220, Dark Matter in
  Galaxies. p.~103 (\mn@eprint {} {astro-ph/0309163})

\bibitem[\protect\citeauthoryear{{Schechter}, {Morgan}, {Chehade}, {Metcalfe},
  {Shanks}  \& {McDonald}}{{Schechter} et~al.}{2017}]{schechter2017}
{Schechter} P.~L.,  {Morgan} N.~D.,  {Chehade} B.,  {Metcalfe} N.,  {Shanks}
  T.,   {McDonald} M.,  2017, \mn@doi [\aj] {10.3847/1538-3881/aa6899}, \href
  {http://adsabs.harvard.edu/abs/2017AJ....153..219S} {153, 219}

\bibitem[\protect\citeauthoryear{{Secrest}, {Dudik}, {Dorland}, {Zacharias},
  {Makarov}, {Fey}, {Frouard}  \& {Finch}}{{Secrest}
  et~al.}{2015}]{secrest2015}
{Secrest} N.~J.,  {Dudik} R.~P.,  {Dorland} B.~N.,  {Zacharias} N.,  {Makarov}
  V.,  {Fey} A.,  {Frouard} J.,   {Finch} C.,  2015, \mn@doi [\apjs]
  {10.1088/0067-0049/221/1/12}, \href
  {http://adsabs.harvard.edu/abs/2015ApJS..221...12S} {221, 12}

\bibitem[\protect\citeauthoryear{{S{\'e}rsic}}{{S{\'e}rsic}}{1963}]{sersic1963}
{S{\'e}rsic} J.~L.,  1963, Boletin de la Asociacion Argentina de Astronomia La
  Plata Argentina, \href {http://adsabs.harvard.edu/abs/1963BAAA....6...41S}
  {6, 41}

\bibitem[\protect\citeauthoryear{{Shanks} et~al.,}{{Shanks}
  et~al.}{2015}]{shanks2015}
{Shanks} T.,  et~al., 2015, \mn@doi [\mnras] {10.1093/mnras/stv1130}, \href
  {http://adsabs.harvard.edu/abs/2015MNRAS.451.4238S} {451, 4238}

\bibitem[\protect\citeauthoryear{{Stern} et~al.,}{{Stern}
  et~al.}{2012}]{stern2012}
{Stern} D.,  et~al., 2012, \mn@doi [\apj] {10.1088/0004-637X/753/1/30}, \href
  {http://adsabs.harvard.edu/abs/2012ApJ...753...30S} {753, 30}

\bibitem[\protect\citeauthoryear{{The Dark Energy Survey Collaboration}}{{The
  Dark Energy Survey Collaboration}}{2005}]{despaper}
{The Dark Energy Survey Collaboration} 2005, ArXiv Astrophysics e-prints, \href
  {http://adsabs.harvard.edu/abs/2005astro.ph.10346T} {}

\bibitem[\protect\citeauthoryear{{Treu} \& {Marshall}}{{Treu} \&
  {Marshall}}{2016}]{treu2016}
{Treu} T.,  {Marshall} P.~J.,  2016, \mn@doi [\aapr]
  {10.1007/s00159-016-0096-8}, \href
  {http://adsabs.harvard.edu/abs/2016A%26ARv..24...11T} {24, 11}

\bibitem[\protect\citeauthoryear{Williams et~al.,}{Williams
  et~al.}{2018}]{williams2017}
Williams P.~R.,  et~al., 2018, \mn@doi [Monthly Notices of the Royal
  Astronomical Society: Letters] {10.1093/mnrasl/sly043}, p. sly043

\bibitem[\protect\citeauthoryear{{Wright} et~al.,}{{Wright}
  et~al.}{2010}]{wright2010}
{Wright} E.~L.,  et~al., 2010, \mn@doi [\aj] {10.1088/0004-6256/140/6/1868},
  140, 1868

\bibitem[\protect\citeauthoryear{{Wu}, {Hao}, {Jia}, {Zhang}  \& {Peng}}{{Wu}
  et~al.}{2012}]{wu2012}
{Wu} X.-B.,  {Hao} G.,  {Jia} Z.,  {Zhang} Y.,   {Peng} N.,  2012, \mn@doi
  [\aj] {10.1088/0004-6256/144/2/49}, \href
  {http://adsabs.harvard.edu/abs/2012AJ....144...49W} {144, 49}

\makeatother
\end{thebibliography}





\bsp	
\label{lastpage}
\end{document}